\documentclass[11pt]{article}
\pdfoutput=1 
\usepackage{./jheppub}

\usepackage[T1]{fontenc}

\usepackage{graphicx}
\usepackage{amsfonts,amsmath,amssymb}
\usepackage{verbatim}
\usepackage{array}
\usepackage{mathrsfs}
\usepackage{mathrsfs}
\usepackage{yfonts}
\usepackage{dsfont}
\usepackage{bbm}
\usepackage{colonequals}
\usepackage{amscd}

\usepackage{subcaption}

\usepackage{wrapfig}

\usepackage{suffix,mathtools,cancel,bbm}

\usepackage{multirow}

\usepackage{enumerate}

	\usepackage{tikz, tikz-3dplot, pgfplots}
	\usepackage{tkz-graph}
	\usetikzlibrary[positioning,patterns] 
	\usepackage{genyoungtabtikz}

\newcommand{\bn}{\begin{enumerate}}
\newcommand{\en}{\end{enumerate}}

\newcommand{\beq}{\begin{equation}}
\newcommand{\eeq}{\end{equation}}

\newcommand\nn{\nonumber}

\newcommand{\cA}{\mathcal{A}}

\newcommand{\cC}{\mathcal{C}}

\newcommand{\cI}{\mathcal{I}}

\newcommand{\cN}{\mathcal{N}}

\newcommand{\cS}{\mathcal{S}}

\newcommand{\cV}{\mathcal{V}}

\numberwithin{equation}{section}

\def\bea{\begin{eqnarray}}
\def\eea{\end{eqnarray}}

\DeclarePairedDelimiterX\MeijerM[3]{\lparen}{\rparen}%
{\begin{smallmatrix}#1 \\ #2\end{smallmatrix}\delimsize\vert\,#3}

\newcommand\MeijerG[8][]{%
  G^{\,#2,#3}_{#4,#5}\MeijerM[#1]{#6}{#7}{#8}}

\WithSuffix\newcommand\MeijerG*[7]{%
  G^{\,#1,#2}_{#3,#4}\MeijerM*{#5}{#6}{#7}}

\def\cN{\mathcal{N}}

\def \beg#1{\begin{#1}} 

\def \bea{\beg{eqnarray}}
\def \eea{\end{eqnarray}}
\def \ee{\end{equation}}

\def \restr#1#2{{\left.\kern-\nulldelimiterspace#1\vphantom{\big|}\right|_{#2}}}

\def \nn{\nonumber}

\usepackage{colortbl}
\definecolor{mygray}{gray}{0.93}

\title{\boldmath Anomaly Inflow, Accidental Symmetry,
and Spontaneous Symmetry Breaking}

\author[a]{Ibrahima Bah,}
\author[a]{Federico Bonetti,}

\affiliation[a]{Department of Physics and Astronomy, Johns Hopkins University, 3400 North Charles Street, Baltimore, MD 21218, USA}

\emailAdd{iboubah@jhu.edu, fbonett3@jhu.edu}

\abstract
{
We consider the 6d (1,0) SCFT on a stack of $N$ M5-branes
probing a $\mathbb C^2/\mathbb Z_2$ singularity.
In particular, we study its compactifications to four dimensions
on a smooth genus-$g$ Riemann surface
with non-trivial flavor flux, yielding a
family of 4d CFTs.
By tracking the M-theory origin of the global symmetries
of the 4d CFTs, we detect the emergence of an accidental
symmetry and the spontaneous symmetry
breaking of a $U(1)$ generator.  These effects are visible from geometric considerations
and not apparent from the point of view of the compactification of the 6d field theory.
These phenomena leave an imprint on the 't Hooft anomaly
polynomial of the 4d CFTs, which is obtained from 
recently developed anomaly inflow methods in M-theory \cite{Bah:2019rgq}.
In the large-$N$ limit, we identify the gravity dual
of the 4d setups to be a class of smooth $AdS_5$ solutions
first discussed by Gauntlett-Martelli-Sparks-Waldram.
Using our anomaly polynomial, we compute the conformal central charge and
a non-Abelian flavor central charge at large $N$,
finding agreement with the holographic predictions.
}

\usepackage{etoolbox}
\appto\appendix{\addtocontents{toc}{\protect\setcounter{tocdepth}{1}}}

\appto\listoffigures{\addtocontents{lof}{\protect\setcounter{tocdepth}{1}}}
\appto\listoftables{\addtocontents{lot}{\protect\setcounter{tocdepth}{1}}}

\begin{document} 

\maketitle
\flushbottom



\section{Introduction and summary}

Quantum field theory (QFT) provides a powerful framework 
to describe a variety of physical phenomena, ranging from
particle physics, to condensed matter systems and cosmology.
Symmetries and spontaneous symmetry breaking 
play a fundamental role in countless examples
of applications of the QFT formalism.
It is particularly interesting to investigate the symmetries and
dynamics of 
QFTs
in strongly coupled non-perturbative regimes.
Geometric engineering is a remarkable tool in the 
construction and analysis of strongly coupled 
QFTs 
 in various dimensions. 
Several non-trivial QFTs can be studied by examining   the low-energy
limit of brane configurations  in
string theory and M-theory.
A prominent example is furnished by 
6d (2,0) theories of type $A_{N-1}$,
which emerge in the long-wavelength dynamics of a stack
of $N$ M5-branes extending along
 a flat worldvolume \cite{Witten:1995zh,Strominger:1995ac}. 
By a similar token, an interesting class of 6d
(1,0) theories is obtained by considering a stack of M5-branes
probing an orbifold singularity 
\cite{Brunner:1997gk,Blum:1997fw,Blum:1997mm,Intriligator:1997dh,Brunner:1997gf,Hanany:1997gh}.
A rich variety of 4d QFTs can be constructed 
by considering M-theory setups in which
a stack of M5-branes is wrapped on a Riemann surface.
These 4d QFTs are generically strongly coupled and fit
into the larger Class $\cS$ program,
in which 
6d superconformal field theories (SCFTs)
are compactified to four dimensions on a Riemann surface,
possibly with defects.
The reduction of 6d (2,0) theories 
to 4d $\cN = 2$ QFTs was first analyzed in \cite{Gaiotto:2009we,Gaiotto:2009hg},
and reduction to 4d $\cN = 1$ QFTs has been studied in \cite{Maruyoshi:2009uk,Benini:2009mz,Bah:2011je,Bah:2011vv,Bah:2012dg}.
The compactification of 6d (1,0) theories has been addressed in 
\cite{Gaiotto:2015usa,Ohmori:2015pua,DelZotto:2015rca,Ohmori:2015pia,Razamat:2016dpl,Bah:2017gph,Kim:2017toz,Kim:2018bpg,Kim:2018lfo,Razamat:2018gro,Zafrir:2018hkr,Ohmori:2018ona,Chen:2019njf,Razamat:2019mdt,Pasquetti:2019hxf,Razamat:2019ukg}.

't Hooft anomalies are among the most important observables 
to compute in a geometrically engineered QFT, especially
if a Lagrangian description of the theory is not available.
It is worth emphasizing that anomalies are naturally
geometric quantities. For the case of 
 continuous 0-form symmetries---which is the 
case relevant for this work---the anomalies 
of a $d$-dimensional QFT (with $d$ even) are   encoded
in the anomaly polynomial, which is a $(d+2)$-form
constructed with the curvatures of the background fields
associated to the symmetries \cite{AlvarezGaume:1983ig,AlvarezGaume:1984dr,Bardeen:1984pm}.
The geometric nature of 't Hooft anomalies 
makes them particularly amenable to computation
in the framework of geometric engineering.
Building on seminal papers on anomaly inflow for M5-branes 
\cite{Duff:1995wd,Witten:1996hc,Freed:1998tg,Harvey:1998bx},
a systematic toolkit for the computation of anomalies of QFTs 
from M5-branes has been developed in \cite{Bah:2018gwc, Bah:2018jrv, Bah:2019jts, Bah:2019rgq}.

By investigating 't Hooft anomalies 
we can cast light on other interesting physical phenomena.
To illustrate this point, and to exemplify the
power of geometric methods in QFT,
we consider   a class of 4d $\cN = 1$ theories,
obtained from compactification on a Riemann surface
of the worldvolume theory of a stack of M5-branes
probing a $\mathbb C^2 /\mathbb Z_2$ singularity. 
In the reduction from six to four dimensions,
we encounter accidental global symmetries
as well as spontaneous symmetry breaking.
Both features can be detected 
by a careful analysis of the M-theory origin
of the global symmetries of the 4d QFT.

In order to describe in more detail the class
of 4d theories we study in this work,
let us recall some salient features of 
 the 6d (1,0) SCFT 
on a stack of $N$
M5-branes
probing  a $\mathbb C^2/\mathbb Z_2$ singularity.
Before modding out by $\mathbb Z_2$,
the stack is surrounded in its five transvervse
directions by a 4-sphere $S^4$.
After quotienting by $\mathbb Z_2$, $S^4$
is replaced by $S^4/\mathbb Z_2$.
The $\mathbb Z_2$ action has two fixed points,
 located at 
the north and south poles of $S^4$,
which yield two orbifold singularities on $S^4/\mathbb Z_2$.
The theory has global symmetry 
$SU(2)_L \times SU(2)_R \times SU(2)_\mathrm N \times SU(2)_\mathrm S$.
The factors $SU(2)_L \times SU(2)_R$ originate
from isometries of $S^4/\mathbb Z_2$,
while $SU(2)_\mathrm N \times SU(2)_\mathrm S$
originate from the two orbifold points
(labeled N, S for ``north'', ``south'').
The factor $SU(2)_R$ is the 6d R-symmetry,
while the other factors are flavor symmetries.

\begin{figure}
\centering
\includegraphics[width = 12.5cm]{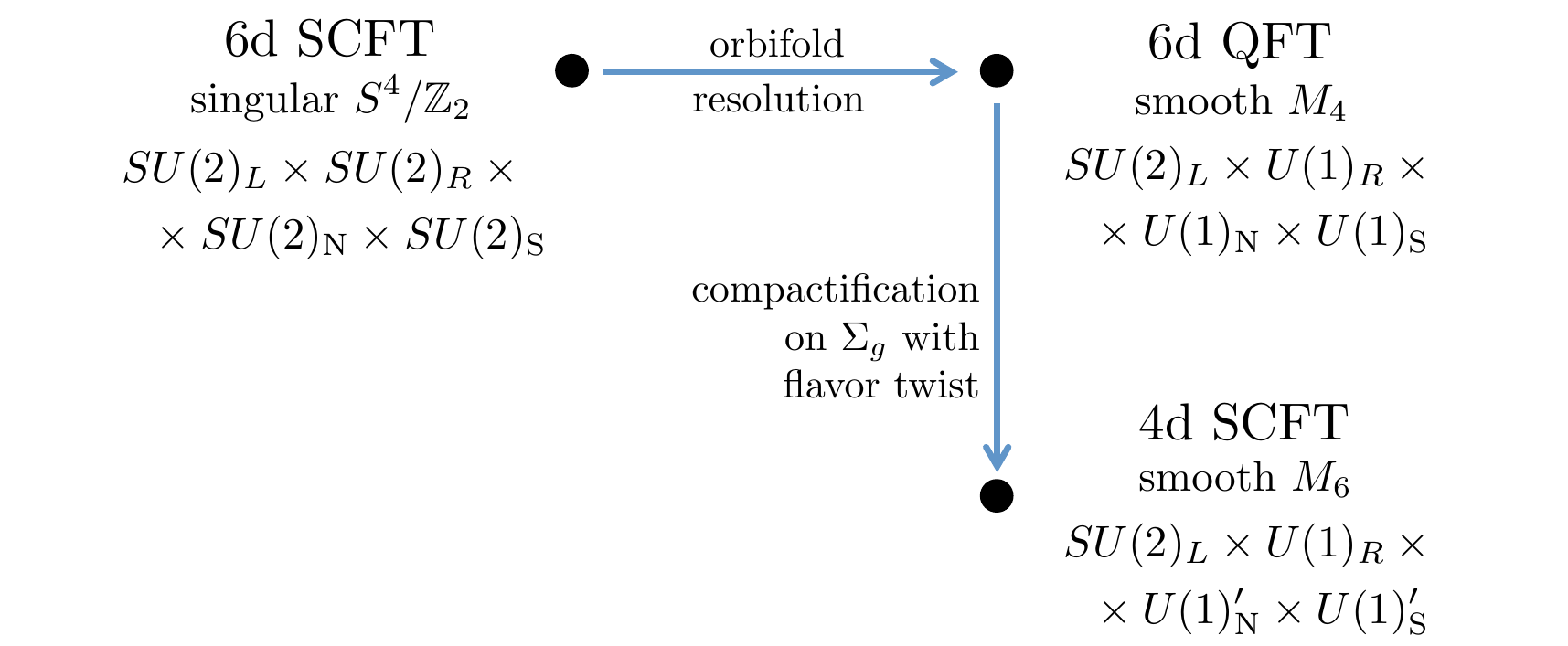}
\caption{Schematic representation of the construction
considered in this work.
The starting point is the 6d (1,0) SCFT living on a stack of M5-branes
probing the singularity $\mathbb C^2/\mathbb Z_2$.
We move away from the 6d conformal point by
resolving the two orbifolds points of $S^4/\mathbb Z_2$ while
preserving 6d  (1,0) supersymmetry.
Finally, we compactify on the Riemann surface $\Sigma_g$
with a suitable flavor twist. 
For each theory we give the associated internal space
and the global symmetry group.
The factors $U(1)'_\mathrm N \times U(1)'_\mathrm S$
are defined in \eqref{intro_eq}.
}
\label{figure_intro}
\end{figure}

The orbifold singularities at the north and south poles
can be resolved in a canonical
way preserving 6d (1,0) supersymmetry.
The orbifold $S^4/\mathbb Z_2$ is replaced
by a smooth internal space $M_4$.
In the resolved phase, the flavor symmetry
$SU(2)_\mathrm N \times SU(2)_\mathrm S$
is broken to its Cartan subgroup
$U(1)_\mathrm N \times U(1)_\mathrm S$.
The 4d theories of interest in this work are obtained by reducing the 6d theory
in its resolved phase by compactification on a smooth genus-$g$ Riemann surface
$\Sigma_g$ (with $g \neq 1$). The compactification
includes a twist for the $U(1)_\mathrm N \times U(1)_\mathrm S$ flavor
symmetry, while $SU(2)_L$ is untwisted.
The resulting 4d theories are then labelled
by the genus $g$ and 
 three flux quanta $N$, $N_\mathrm N$, $N_\mathrm S$,
with $N$ the number of M5-branes in the stack,
and $N_{\rm N,S}$ the twist parameter for the flavor
symmetry $U(1)_{\rm N,S}$.
The internal space $M_4$ of the resolved 6d theory
is non-trivially fibered over the Riemann surface,
yielding a 6d space $M_6$, with $M_4 \hookrightarrow M_6 \rightarrow \Sigma_g$.
Our construction is summarized schematically in figure \ref{figure_intro}.

The global symmetries of the 4d QFT
are encoded in the geometry and topology of the space
$M_6$. We notice that the global symmetries 
of the 4d QFT correspond to gauge symmetries
of the 5d supergravity obtained by reduction of M-theory
on the internal space $M_6$.
In this 5d supergravity theory, we have a massless gauge field
for each isometry generator of $M_6$.
Furthermore, additional 5d gauge fields
are obtained by expanding the M-theory 3-form $C_3$
onto harmonic 2-forms in $M_6$.

The space $M_6$
has isometry group $U(1)_R \times SU(2)_L$,
which is simply identified with the subgroup
of the  6d isometry group
$SU(2)_R \times SU(2)_L$ that is preserved
by twisting over the Riemann surface.
The space $M_6$ also admits three harmonic 2-forms,
denoted $\omega_\mathrm N$, $\omega_\mathrm S$,
$\omega_\mathrm C$.
The 2-forms $\omega_{\rm N,S}$ can be traced
back to the resolution of the orbifold singularities 
of $S^4/\mathbb Z_2$ of the parent 6d theory.
The 2-form $\omega_\mathrm C$, on the other hand,
only emerges after reduction to four dimensions,
by fibering $M_4$ non-trivially over $\Sigma_g$.
From a field theory point of view,
the existence of a third harmonic 2-form in the 4d setup
is interpreted as the emergence of an accidental $U(1)_C$ global symmetry,
which is not visible in six dimensions.

A crucial feature of the 5d supergravity theory
obtained from compactification on $M_6$ is the following.
While all three harmonic 2-forms 
$\omega_\mathrm N$, $\omega_\mathrm S$,
$\omega_\mathrm C$ yield a 5d vector,
one linear combination of such vectors gets massive
via St\"uckelberg mechanism,
by coupling to a 5d axion. This phenomenon
is described in greater detail in \cite{Bah:2019rgq}.\footnote{This
St\"uckelberg mechanism was also instrumental in 
\cite{Gaiotto:2009gz} for the correct counting of symmetries
of 4d $\cN = 2$ SCFTs from M5-branes from the perspective of their gravity duals.}
From a field theory perspective,
the symmetry group
$U(1)_\mathrm N \times U(1)_\mathrm S  \times U(1)_\mathrm C$
is spontaneouly broken
to a $U(1)'_\mathrm N \times U(1)'_\mathrm S$ subgroup.
The connection between 
 St\"uckelberg mechanism in 5d supergravity,
 and spontaneous symmetry breaking
of global symmetries in the 4d theory,
is well-established in the holography
literature, see \emph{e.g.}~\cite{Gubser:2008px, Hartnoll:2008vx}.
To summarize,
\begin{gather} 
U(1)_\mathrm N \times U(1)_\mathrm S \; \xrightarrow[\;\;\;\;\phantom{l} \text{symmetry}\phantom{l}\;\;\;\;] 
{\phantom{p}\text{accidental}\phantom{p}} \;
U(1)_\mathrm N \times U(1)_\mathrm S  \times U(1)_\mathrm C \;
\xrightarrow[\text{symm.~breaking}] 
{\text{spontaneous}} \;
U(1)'_\mathrm N \times U(1)'_\mathrm S \ , \nn
\end{gather}
with the generators $T'_{\rm N,S}$ of $U(1)'_{\rm N,S}$
given in terms of the generators $T_{\rm N,S,C}$ of $U(1)_{\rm N,S,C}$
as
\beq \label{intro_eq}
T'_\mathrm N = T_\mathrm N - \frac{N_\mathrm N}{N} \, T_\mathrm C \ , \qquad
T'_\mathrm S = T_\mathrm S - \frac{N_\mathrm S}{N} \, T_\mathrm C \ .
\eeq
The na\"ive symmetry $U(1)_\mathrm N \times U(1)_\mathrm S$
visible in 6d dimensions is replaced
by $U(1)'_\mathrm N \times U(1)'_\mathrm S$.
Even though the rank is unchanged,
this process has deep implications for the
't Hooft anomalies of the theory, due to the non-trivial
mixing of generators
in \eqref{intro_eq}.

We perform a careful analysis of the 't Hooft anomalies
of the 4d QFT,
using the techniques developed in \cite{Bah:2019rgq},
based on anomaly inflow from the M-theory ambient space.
The main idea in \cite{Bah:2019rgq} is to obtain the inflow anomaly
polynomial $I_6^{\rm inflow}$ of the 4d QFT by integrating 
a 12-form characteristic class $\cI_{12}$,
which encodes the anomalous variation
of the M-theory action in the presence of the M5-brane stack.
Crucially,  $I_6^{\rm inflow}$ counterbalances the 
anomalies of all degrees of freedom living on the stack,
which in the IR can be organized into the interacting QFT of interest,
plus possible decoupled sectors.
We may then write
\beq \label{inflow_eq}
I_6^{\rm inflow} = \int_{M_6} \cI_{12} \ , \qquad
I_6^{\rm inflow} + I_6^{\rm QFT} + I_6^{\rm decoupl} = 0 \ .
\eeq
While we do not have a complete understanding of
decoupled sectors, we can assume that their
contribution to the 't Hooft anomalies
is subleading in the large-$N$ limit,
which is taken as $N, N_{\rm N,S} \rightarrow \infty$,
keeping $N_{\rm N,S}/N$ finite.
Under this assumption, 
the term $ I_6^{\rm decoupl} $ in \eqref{inflow_eq} can be neglected
at leading order at large $N$, and we can infer the anomaly
polynomial of the interacting QFT from $I_6^{\rm inflow}$.

In order to test our large-$N$ result,
we investigate the gravity duals of our 4d field theory constructions.
As it turns out, we identify the gravity duals to be a 
well-known class of $AdS_5$ solutions in M-theory,
first discussed in Gauntlett-Martelli-Sparks-Waldram (GMSW) \cite{Gauntlett:2004zh}.
These solutions are warped products $AdS_5 \times_w M_6^{\rm hol}$,
where the internal space $M_6^{\rm hol}$ is smooth
and has exactly the same topology and isometries as $M_6$
in the probe M5-brane picture.
The existence of smooth dual geometries
provides evidence that our construction yields
a non-trivial interacting 4d SCFT in the IR,
at least at large $N$.

To give more supporting evidence for our claims,
we carry out two quantitative checks,
by computing the central charge $c$
and the flavor central charge for the global
symmetry $SU(2)_L$ of the 4d theory.
These quantities can be computed holographically at large $N$
from the supergravity effective action.
On the field theory side, they can be extracted from the anomaly polynomial,
because the superconformal algebra relates
them to 't Hooft anomalies coefficients involving the
$SU(2)_L$ generators and the superconformal
R-symmetry \cite{Anselmi:1996dd,Anselmi:1997rd}. The latter is the linear combination of $U(1)_R \times U(1)'_\mathrm N \times U(1)'_\mathrm S$ determined by $a$-maximization \cite{Intriligator:2003jj}.
We find a perfect agreement between the supergravity
and field theory computations, at leading order in the large-$N$
expansion.

The outcome of $a$-maximization
depends crucially on the mixing \eqref{intro_eq} of the na\"ive
 symmetry generators
 $T_{\rm N,S}$ of $U(1)_{\rm N,S}$
with the generator $T_\mathrm C$ of the emergent  
$U(1)_\mathrm C$ symmetry.
In particular, in order to match the supergravity results
it is essential to take into account emergent symmetries and
spontaneous symmetry breaking in the computation of the
't Hooft anomalies of the 4d theory. 
The methods of \cite{Bah:2019rgq} provide a streamlined, geometric way
of addressing these phenomena.
Indeed, if we     
integrate the anomaly polynomial of the parent 6d SCFT
of $\Sigma_g$, we do not reproduce the correct large-$N$
central charge from holography.

The rest of this paper is organized as follows.
In section \ref{6d_section} we describe the 6d M-theory setup
with a stack of $N$ M5-branes probing a $\mathbb C^2/\mathbb Z_2$ singularity.
Section \ref{4dsection} is devoted to the reduction  
of the 6d theory to four dimensions on $\Sigma_g$,
and to the global symmetries of the 4d QFT.
In section \ref{4d_anomalies} we compute the anomalies
of the 4d QFT using inflow.
In section \ref{holo_section} we identify the gravity duals
and we perform the aforementioned quantitative tests
involving the central charges of the 4d theories.
We conclude with a brief discussion in section \ref{discussion_sec}.
The computations in supergravity are collected in appendix \ref{sugra_app}.

\section{Six-dimensional setup} \label{6d_section}

The main setup of interest is a stack of $N$ M5-branes probing a $\mathbb{C}^2/\mathbb{Z}_2$ orbifold singularity in M-theory.  First, we discuss some general aspects where the M5-branes are probing a $\mathbb{C}^2/\mathbb{Z}_k$ orbifold fixed point.

\subsection{Aspects of M-theory on $\mathbb{C}^2/\mathbb{Z}_k$}

First we consider the M-theory background with the orbifold $\mathbb{C}^2/\mathbb{Z}_k$.  Let $(x^0,\cdots,x^5)$ be the coordinates along an $\mathbb{R}^6$ plane, and $(y^1,\cdots,y^5)$ be the coordinates along the transverse directions.  The latter parametrize a five-dimensional space $\mathbb{C}^2 \times \mathbb{R}$ with complex coordinates $(z_1= y^1+ i y^2, z_2 = y^3+iy^4)$.  The orbifold action is 
\begin{equation}
(z_1,z_2, y^5) \sim (e^{\frac{2\pi i}k} z_1,e^{-\frac{2\pi i}k}z_2, y^5).  
\end{equation} A local metric for the M-theory background is given as
\begin{align}
ds^2_{11} &= ds^2(\mathbb{R}^6) +  dr^2 + r^2 \, ds^2 (S^4/\mathbb{Z}_k)   \label{kbackground} \\
ds^2 (S^4/\mathbb{Z}_k)  &= \frac{d\mu^2}{1-\mu^2} + (1-\mu^2) \left[ \frac{1}{k^2} \,
D\varphi^2 + \frac{1}{4} \, ds^2(S^2_\psi) \right], \label{sphereorb}\\
ds^2(S^2_\psi) &= d\theta^2 + \sin^2 \theta \,   d\psi^2, \qquad D\varphi = d\varphi + \frac{k}{2} \cos \theta \,  d\psi.  
\end{align} The radius $r$ is constructed from $y^5$ and the radii of the two complex planes, in particular we have $y^5 = r \mu$.  The circle coordinates have periodicity $(\Delta \psi = 2\pi, \Delta \varphi = 2\pi)$.  For $k=1$, the metric \eqref{sphereorb} is that of round four-sphere.  When $k> 1$, the orbifold action admits two $\mathbb{R}^4/\mathbb{Z}_{k}$ fixed points at the poles of the sphere at $\mu = \pm 1$.  At constant values of $\mu$, the three dimensional sections of the four-sphere are $S^1_{\phi}$ bundles over $S^2_\psi$ with degree $k$.  The isometries of $S^1_\varphi$ and $S^2_\psi$ lead to a $U(1)_\varphi \times SU(2)_\psi$ gauge symmetry on the extended seven-dimensional directions of the M-theory background.  This is the subgroup of the $SO(5)$ isometry of the sphere preserved by the orbifold action.  

The region near an orbifold fixed point of the sphere corresponds to a single center Taub-NUT space, the metric near each pole is 
\begin{equation}
ds^2  \cong \frac{1}{V} \,  D\varphi^2 + V \,  \left[d R^2 + R^2 \,  ds^2(S^2_\psi) \right] \ , \quad V = \frac{k}{2R} \ , \quad R = \frac{1}{k}  \, (1\mp \mu) \ , \quad \mbox{for} \quad \mu = \pm 1 \ . \label{polesmetric}
\end{equation} The orbifold singularities can be resolved locally by replacing the single center Taub-NUT space to a Gibbons-Hawking space with $k$ sources of unit charge.  Such spaces are $S^1_\varphi$ bundles over $\mathbb{R}^3$ with metric given as 
\begin{equation}
ds^2 = V^{-1} \left(d\varphi + A \right)^2  + V \left(dX^2 + dY^2 + dZ^2\right) \ , \qquad \mbox{with} \quad d V = \star_{\mathbb R^3} dA \ ,
\end{equation} where $V$ is a potential on the 3D base space and $A$ is a connection one-form for the circle bundle.  The potential satisfies Laplace's equation on the $\mathbb{R}^3$ base.  A general solution of Gibbons-Hawking space is given by inserting $k$ centers at positions $\vec{X}_I = (X_I,Y_I,Z_I)$ with charge $n^I$.  The potential is
\begin{equation}
V = v_0 + \frac{1}{2} \sum_I \frac{n^I}{ \big |\vec{X} -\vec{X}_I  \big |} \ .
\end{equation} The parameter $v_0$ fixes the asymptotic size of the circle.  The coordinate $\varphi$ has period $2\pi$.  The space is Asymptotically Locally Flat (ALF) when $v_0\neq 0$ with topology of $S^1 \times \mathbb{R}^3$, and asymptotically Locally Euclidean (ALE) when $v_0 =0$ with topology of $\mathbb{R}^4$.

The region near each center is described by a $\mathbb{R}^4/\mathbb{Z}_{n^I}$ orbifold fixed point where $S^1_\varphi$ shrinks.  A two-cycle can be obtained by taking $S^1_\varphi$ with a segment on $\mathbb{R}^3$ that connects two singularities.  There are $k-1$ independent two-cycles with harmonic representatives, $\omega^i$ $(i\in \{1,\cdots, k-1\})$.  When all $n^I=1$, the space is smooth with $k-1$ two-cycles.  This corresponds to a resolution of the orbifold singularity.  The $S^1_\psi$ circle of $S^2_\psi$ in \ref{polesmetric}, is identified with the rotation on the $(X,Y)$ plane.  Resolving the singularity breaks the $SU(2)_\psi$ isometry of $S^2_\psi$ to a $U(1)_\psi$ corresponding to the rotations of $S^1_\psi$.  

M-theory, in the supergravity limit, can be studied on the space \eqref{kbackground} where we replace the two orbifold singularities with their smooth resolutions.  This local deformation is always possible since the asymptotic space of Gibbons-Hawking is fixed by the total charge.  M-theory has a 2-form gauge symmetry with a 3-form potential, $C_3$.  There are massless fluctuations of the $C_3$ potential on the extended seven-dimensional space coming from the resolutions of the singularities.  In reducing M-theory on the compact space, we can add terms to $C_3$ as 
\begin{equation}
C_3 = a_{ i}^\mathrm N \wedge \omega^i_{\mathrm N} 
+  a_{ i}^\mathrm S \wedge \omega^i _{\mathrm S}+ \cdots \label{C36d}
\end{equation} where $(\omega_\mathrm N^i,\omega_\mathrm S^i)$ are the resolution harmonic two-forms for the north and south singularities, respectively.  The one-forms $(a^\mathrm N_i,a^\mathrm S_i)$ are massless gauge field on the seven-dimensional space.  Each orbifold fixed point leads to a $U(1)^{k-1}$ gauge symmetry in M-theory which enhances to an $SU(k)$ gauge symmetry in the singular limit \cite{Sen:1997kz}.  To summarize, the M-theory background admits an $SU(2)_\psi \times SU(k)_\mathrm N \times SU(k)_\mathrm S \times U(1)_\varphi$ bosonic gauge symmetry.

We are interested in the field theory that describe the low energy dynamics of a stack of $N$ M5-branes probing the M-theory singularity at $r=0$ in \eqref{kbackground}.  The gauge symmetry of the M-theory background induces a global symmetry on the worldvolume directions of the M5-branes.  The $SU(2)_\psi$ symmetry from $S^2_\psi$ corresponds to an $SU(2)_\mathcal{R}$ R-symmetry, whiles the rest $SU(k)_\mathrm N \times SU(k)_\mathrm S \times U(1)_\varphi$ imprints as a flavor symmetry. In particular,
the 7d gauge fields $(a^\mathrm N_i , a^\mathrm S_i)$
yield 6d background connections $(A^\mathrm N_{6d \, i} , A^\mathrm S_{6d \, i})$. This configuration preserves eight supercharges leading to a six-dimensional superconformal field theory with a $(1,0)$ supersymmetry 
 \cite{Brunner:1997gk,Blum:1997fw,Blum:1997mm,Intriligator:1997dh,Brunner:1997gf,Hanany:1997gh}.  

\subsection{The case   $k=2$}\label{k2res}

The resolution space of $S^4/\mathbb{Z}_2$ has an enhanced symmetry.  The $U(1)_\varphi$ isometry of $S^1_\varphi$ in $S^4/\mathbb{Z}_2$
enhances  to an $SU(2)_\varphi$ isometry group as can be seen by rewriting the metric as
\begin{equation}
ds^2(S^4/\mathbb{Z}_2) = \frac{d\mu^2}{1-\mu^2} + \frac{1}{4} \,  (1-\mu^2) \,  \left[ 
D\psi^2 + d\theta^2 + \sin^2 \theta  \,  d\varphi^2 \right] , \quad  D\psi = d\psi + \cos \theta  \,  d\psi. \label{sphere2}
\end{equation}  The resolution of the orbifold singularities at $\mu = \pm 1$ preserve the isometries of $S^2_\varphi$ composed of $(\theta, \psi)$ in the metric of $S^4/\mathbb{Z}_2$ above.  This follows from the fact that the resolution space of $\mathbb{R}^4/\mathbb{Z}_2$ is the Eguchi-Hanson space \cite{Eguchi:1978xp}.  To see this more explicitly, write the potentials for the two center Gibbons-Hawking space with unit charge as
\begin{align}
V &= \frac{1}{2 R_+} + \frac{1}{2 R_-}, \qquad A = \frac{1}{2} \left(\frac{Z_+}{R_+} + \frac{Z_-}{R_-} \right) d \tan^{-1} \left(Y/X \right)  \ , \\
R_\pm^2 &= X^2 + Y^2 + Z_\pm^2 \ , \qquad  Z_\pm = Z \pm Z_0 \ .   
\end{align} The centers are sitting at $(0,0,\pm Z_0)$.  The metric of the resolved space can be written as 
\begin{equation}
ds^2 = \frac{R^4}{R^4-a^4} \, dR^2 + \frac{R^2}{4} \,  \left[d\theta^2 + \sin^2 \theta \,
 d\varphi^2 + \frac{R^4-a^4}{R^4} \left(d\psi + \cos \theta \, d\varphi \right)^2 \right]  \ ,\label{EHspace}
\end{equation} where we have made the coordinate transformation \cite{Prasad:1979kg}
\begin{equation}
Z=\frac{R^2}{4} \, \cos \theta \  , \quad 
X = \frac{\sqrt{R^4-a^4}}{4}\,  \sin \theta  \,  \cos \psi \  , \quad 
Y = \frac{\sqrt{R^4-a^4}}{4} \, \sin \theta  \,  \sin \psi \ .
\end{equation}  
 The space smoothly caps off at $R=a$ where the $S^1_\psi$ circle shrinks.  The two-sphere, $S^2_\varphi$, composed of $(\theta, \varphi)$ in the resolved space has a finite size of $a/2$ at $R=a$.  The two-sphere in this region also corresponds to the two-cycle of the resolution of the $\mathbb{R}^4/\mathbb{Z}_2$.  

The singularities of $S^4/\mathbb{Z}_2$ are resolved by excising the singular region and gluing in the Eguchi-Hanson space describe in \eqref{EHspace}.  The $S^2_\varphi$ of \eqref{EHspace} is identified with the $S^2_\varphi$ of $S^4/\mathbb{Z}_2$ in \eqref{sphere2}.  In this sense, the resolution of the singularities on the sphere preserves its isometries.  

The smooth geometry obtained from the resolution of $S^4/\mathbb{Z}_2$ is denoted as $M_4$.  It has the topology of $S^2 \times S^2$ and corresponds to the Hirzebruch surface $\mathbb{F}_2$.  It is useful to write a local metric for the space $M_4$,
\begin{align}
ds^2(M_4) =  h_1(\mu) \, ds^2(S^2_\varphi) 
+ h_2(\mu) \,  d\mu^2 
+  h_3(\mu)  \, D\psi^2 \  , \qquad 
D\psi &= d\psi + \cos \theta  \,  d\varphi \  . \label{M4metric}
\end{align} The coordinate $\mu$ takes value in the interval $[\mu_\mathrm S, \mu_\mathrm N]$.  The boundary conditions of the $h$'s are fixed by regularity of the metric at the bounds, and in particular, $h_3$ must vanish on them.  The two-sphere $S^2_\varphi$ is not shrinking at the north and south poles of $M_4$ where the $S^1_\psi$ is shrinking.  The two-spheres at the tip of the $\mu$-interval correspond to two-cycles denoted as $(\mathcal{S}^2_\mathrm S, \mathcal{S}^2_\mathrm N)$ respectively.  The volume of these cycles are then $(h_1(\mu_\mathrm S),h_1(\mu_\mathrm N))$ respectively.  


The sizes of the two-cycles $(\mathcal{S}^2_\mathrm S, \mathcal{S}^2_\mathrm N)$ are moduli parameters of $M_4$.  In the singular limit where they vanish, $M_4 \to S^4/\mathbb{Z}_2$.  The four-sphere admits a left $SU(2)$ and a right $SU(2)$ action; these are preserved by the $\mathbb{Z}_2$ orbifold action, and are related to the isometries of $SU(2)_\psi \times SU(2)_\varphi$ rotations.  
The $U(1)_\psi \times SU(2)_\varphi$ are manifest as isometries of $S^1_\psi \times S^2_\varphi$ of the metric \eqref{sphere2}.  The total gauge symmetry in seven dimensions is $SU(2)_\psi \times SU(2)_\varphi \times
SU(2)_\mathrm N \times SU(2)_\mathrm S$. 

\paragraph{Anomalies of the 6d setup}

Anomaly inflow for flat M5-branes probing orbifold singularities was studied in \cite{Ohmori:2014kda}.
See also \cite{Bah:2019rgq} for a review of the computation.
The inflow computation yields an 8-form $I_8^{\rm inflow}$
which captures the variation of the M-theory action in the 
presence of the M5-brane stack. 
The inflow anomaly polynomial
counterbalances the anomalies of the worldvolume degrees of freedom,
which at low energies consists of an interacting SCFT
and of modes that decouple in the IR,
\beq
I_8^{\rm inflow} + I_8^{\rm SCFT} + I_8^{\rm decoupl} = 0 \ .
\eeq
The inflow anomaly polynomial reads
\begin{align}  \label{6d_inflow}
- I_8^{\rm inflow} & = 
 \frac{N^3 }{6} \, \big[ c_2^\varphi - c_2^\psi  \big]^2 
 +  \frac{N }{12}  \, 
 \big[ 4 \, c_2^\varphi
- 3 \, c_2^\psi \big] \, c_2^\psi 
+ \frac{N}{24} \, 
 \big[ 2 \, c_2^\varphi
-  \, c_2^\psi \big] \, p_1(TW_6)
\nn \\
&- \frac{1}{32 \, N} \, 
\bigg[ 
  \frac{ {\rm tr} \, (F_{6d}^{\rm N})^2}{(2\pi)^2}
-  \frac{{\rm tr} \, (F_{6d}^{\rm S})^2}{(2\pi)^2}
\bigg] ^2
   + \frac{N }{4} \, 
\big[ c_2^\varphi - c_2^\psi  \big] \, \bigg[ 
  \frac{ {\rm tr} \, (F_{6d}^{\rm N})^2}{(2\pi)^2}
+  \frac{{\rm tr} \, (F_{6d}^{\rm S})^2}{(2\pi)^2}
\bigg]
\nn \\
&  +  
 \frac{N}{192} \big[  p_1(TW_6)^2 - 4 \, p_2(TW_6)  \big]
 \ .  
\end{align}
We have introduced the compact notation
\beq
c_2^\varphi \equiv c_2(SU(2)_\varphi) \ , \qquad
c_2^\psi \equiv c_2(SU(2)_\psi)  \ .
\eeq
The anomaly of the decoupling modes is given by
\begin{align}
I_8^{\rm decoupl} & = 
 I^{\rm tensor}_8 + \frac 12 \,  I_8^{\rm vec, N}
+ \frac 12 \,  I_8^{\rm vec, S} \ .
\end{align}
The 8-form $ I^{\rm tensor}_8$ is the anomaly polynomial
of a   6d (1,0) tensor multiplet,
\beq
I_8^{\rm tensor} = 
\frac{1}{24} \,( c_2^\psi)^2 + \frac{1}{48} \, c_2^\psi \, p_1(TW_6)
+ \frac{23}{5760} \, p_1(TW_6)^2
- \frac{29}{1440} \, p_2(TW_6) \ .
\eeq
The 8-form $I_8^{\rm vec, N}$ is the anomaly polynomial
of a   6d (1,0) vector multiplet of $SU(2)_\mathrm N$,
\begin{align}
I_8^{\rm vec,N} & = - \frac 18 \, (c_2^\psi)^2
- \frac{1}{16} \, c_2^\psi \, p_1(TW_6)
- \frac{1}{1920} \big[  7 \, p_1(TW_6)^2 - 4 \, p_2(TW_6) \big]
\nn \\
& - \frac 12 \, c_2^\psi \, \frac{{\rm tr}  \, (F_{6d}^N)^2}{(2\pi)^2}
- \frac{1}{24} \, p_1(TW_6) \,   \frac{{\rm tr}  \, (F_{6d}^N)^2}{(2\pi)^2}
- \frac {1}{12} \, \bigg[    \frac{{\rm tr} \, (F_{6d}^N)^2}{(2\pi)^2}\bigg]^2  \ .
\end{align}
The quantity $I_8^{\rm vec, S}$  is completely analogous.\footnote{The
trace `tr' is normalized in such a way that,
if $n_{1,2}$ are the Chern roots of an $SU(2)$ bundle  with $n_1 + n_2 = 0$,
then 
${\rm tr} F^2 /(2\pi)^2 = - 2 \, (n_1^2 + n_2^2)$.
Since $c_2(SU(2)) = n_1  n_2$, we have
${\rm tr} F^2 /(2\pi)^2 = 4 \, c_2(SU(2))$. 
}

\section{Four-dimensional setup} \label{4dsection}

In this section we describe aspects of the geometric setup when a stack of $N$ M5-branes wrapping a genus, $g$, Riemann Surface $\Sigma_{g,n}$ with $n$ punctures, probe a $\mathbb{C}^2/\mathbb{Z}_2$ singularity in M-theory.  We will discuss various aspects of the geometric setup in M-theory and the symmetries they induce for the field theory that describe the low-energy dynamics of the branes.  We will then use the geometric set-up and compute the anomaly polynomial of the field theory by using anomaly inflow techniques developed in \cite{Bah:2019rgq}.

\subsection{Geometric setup for 4d systems}  

 We consider an eleven-dimensional background given as
\begin{equation} \label{11d_background}
M_{11}  = W_4 \times \Sigma_{g,n} \times \mathbb{C}^2/\mathbb{Z}_2 \times \mathbb{R}.
\end{equation}  
The M-theory background preserves supersymmetry when the space $\Sigma_{g,n} \times \mathbb{C}^2/\mathbb{Z}_2$ satisfies the Calabi-Yau threefold condition, i.e. the first Chern class of the space must vanish.  The worldvolume of the branes decompose as $W_6 = W_4 \times \Sigma_{g,n}$ where $W_4$ is the external spacetime.  The low-energy dynamics of the branes is captured by a field theory that live on $W_4$.  In the region near the branes, the spacetime decomposes as
\begin{equation} \label{near_the_branes}
M_{11} = \mathbb{R}^+ \times M_{10}, \quad \mbox{where} \quad M_6 \hookrightarrow M_{10} \rightarrow W_4, \quad M_4 \hookrightarrow M_6 \rightarrow \Sigma_{g,n}.  
\end{equation} The line $\mathbb{R}^+$ is the overall radius of the transverse directions of the worldvolume spacetime $W_4$.  The space $M_{10}$ describes the tubular neighborhood of the branes, it also corresponds to an internal boundary of the M-theory spacetime near the branes.  Finally, $M_4$ is the resolution space of $S^4/\mathbb{Z}_2$ describe in section \ref{k2res}.  

The Calabi-Yau condition on the M-theory background is satisfied by twisting the R-symmetry circle $S^1_\psi$ over the Riemann surface.  This twist breaks the six-dimensional $SU(2)_\mathcal{R}$ R-symmetry to a $U(1)_\mathcal{R}$ symmetry for the worldvolume theory on $W_4$.  At the level of the geometry, this is achieved by shifting the connection of $S^1_\psi$ as
\begin{equation} \label{hatDpsi}
D\psi \; \to \; \widehat{D} \psi = d\psi + \cos \theta \,  d\varphi -2\pi \, \chi \,  A_\Sigma \ ,
 \quad \mbox{with} \quad  \int_{\Sigma_{g,n}}  dA_\Sigma = 1 \ , \quad \chi  = -2(g-1) - n \ .  
\end{equation} The Euler characteristic of the Riemann surface is denoted by $\chi$.  The twisting, $A_\Sigma$, is a nontrivial component of the six-dimensional gauging of the R-symmetry of the 6d theory.

In addition to twisting the R-symmetry, a large family of four-dimensional SCFTs can be engineered by turn on background fields along the Riemann surface for the Cartan elements of the $SU(2)_\mathrm N \times SU(2)_\mathrm S \times SU(2)_\varphi$ flavor symmetry. Such background fields break the flavor symmetry of the six-dimensional theory to $U(1)_\mathrm N \times U(1)_\mathrm S \times U(1)_\varphi$ for the four-dimensional theory.  At the level of the M-theory background, this is achieved by turning on the following flux parameters
\begin{align}
d\varphi \to \widehat{D} \varphi &= d\varphi +2\pi \,  z \,  A_\Sigma \\
C_3 \to \widehat{C}_3 &= C_3 + 2\pi \, (N_\mathrm N  \, A_\Sigma \wedge \omega_\mathrm N 
+ N_\mathrm S \,  A_\Sigma \wedge \omega_\mathrm S ) \ . \label{fluxgauge}
\end{align}  The connection forms $(N_\mathrm N \,  A_\Sigma, N_\mathrm S\,  A_\Sigma)$ are background fields for the $U(1)$ gauge fields discussed in \eqref{C36d} in the case of $k=2$.  The two-forms $(\omega_\mathrm N, \omega_\mathrm S)$
are the closed representatives that measure the volumes of the resolution two-cycles $(\mathcal{S}_\mathrm N^2, \mathcal{S}_\mathrm S^2)$ in $M_4$.  For every choice of a Riemann surface, $\Sigma_{g,n}$, there is a family of four-dimensional systems labeled by the three flux parameters $(z, N_\mathrm N, N_\mathrm S)$ and the number of branes, $N$.  

 In this paper, we restrict to four-dimensional theories that preserve the $SU(2)_\varphi$ symmetry, this corresponds to fixing $z=0$.  We will also restricted to cases with no punctures, i.e. $n=0$, and non-vanishing  curvature, $g\neq 0$.  For this family, a local metric for $M_6$ can be written as
\begin{equation}
ds^2(M_6) = h_0(\mu) ds^2(\Sigma_g) +  h_1(\mu) ds^2(S^2_\varphi) + h_2(\mu) d\mu^2 +  h_3(\mu) \widehat{D}\psi^2.  \label{M6met}
\end{equation} The interval $\mu$ takes value in $[\mu_\mathrm S, \mu_\mathrm N ]$ where the endpoints are fixed by the loci where $h_3$ vanish.  The functions $(h_0,h_1)$ parametrize the radii of $\Sigma_g$ and $S^2_\varphi$ in $M_6$, they are non-vanishing on the interval of $\mu$.

At a fixed point on the surface $\Sigma_g$, there is a fiber that is a copy of $M_4$ composed of $S^2_\varphi$, the circle $S^1_\psi$ and the interval $\mu$.  Similarly, we can consider a fixed point on the sphere $S^2_\psi$, there are four-dimensional fibers which are copies of a space $M_4^\Sigma$ which are composed of the Riemann surface $\Sigma_g$, the circle $S^1_\psi$ and the interval $\mu$.  These fibers do not shrink in $M_6$.  

\subsection{Flux quantization and four-cycles of $M_6$}  

One of the main object of interest is the boundary condition for the four-flux $G_4$ when the branes wrap the Riemann surface. 
Following \cite{Freed:1998tg,Harvey:1998bx},
we can write the boundary term for $G_4$ by using a bump function  $\rho(r)$ as
\begin{equation} \label{barG4_def}
G_4 = 2\pi \rho(r) \,  \overline {G}_4 + \dots, \qquad \mbox{with} \qquad \int_{M_4} \overline {G}_4 = N  \ .
\end{equation} The flux $\overline {G}_4$ supports the non-trivial geometry $M_6$.  

The space $M_4$ is the fiber over the Riemann Surface, $\Sigma_g$.  Since it is non-shrinking, we can thread $N$ units of flux on it at a fixed point of the Riemann surface.  Similarly, at a fixed point of the sphere, $S^2_\varphi$, we can thread flux on the fiber $M_4^\Sigma$ given as 
\begin{equation}
\int_{M_4} \overline {G}_4 = N_\Sigma.  
\end{equation}  The space $M_6$ also admits four-cycles localized at the north and south poles of the $\mu$ interval where $S^1_\psi$ shrinks.  These correspond to the product of the resolution cycles of the original orbifold singularities of the $S^4/\mathcal{Z}_2$ with the Riemann surface, $\Sigma_g$.  They are denoted as
\begin{align}
\mathcal{C}_\mathrm N^4 &= \mathcal{S}^2_\mathrm N \times \Sigma_g & &   
\hspace{-3cm} \mbox{localized at} \qquad \mu = \mu_\mathrm N \ , \\
\mathcal{C}_\mathrm S^4 &= \mathcal{S}^2_\mathrm S \times \Sigma_g & &\hspace{-3cm}  \mbox{localized at} \qquad \mu = \mu_\mathrm S \ .
\end{align} We can thread flux on these cycles given as 
 \begin{equation} \label{N_north_south_def}
 \int_{\mathcal{C}_\mathrm N^4} \overline {G}_4 = N_\mathrm N \in \mathbb{Z}, \qquad  \int_{\mathcal{C}_\mathrm S^4} \overline {G}_4 = N_\mathrm S \in \mathbb{Z}.
\end{equation} It seems then that there exist four flux quanta $(N,N_\Sigma,N_\mathrm N,N_\mathrm S)$ that label the class of $\overline {G}_4$ in $M_6$.  These flux parameters are not all independent.  This reflects the fact that the space $M_6$ actually admits only three four-cycles denoted as $(\mathcal{C}_\mathrm N^4, \mathcal{C}_\mathrm S^4,  \mathcal{C}_\mathrm C^4)$.  The last one $\mathcal{C}_\mathrm C^4$ is not localized on $M_6$.  
It consists of $S^2_\varphi \times S^2_{\mu \psi}$ at a generic point
on $\Sigma_g$.

To see that there are only three independent flux parameters, we consider the most general local expression for $\overline {G}_4$ that is closed and consistent with the symmetries of $M_6$
\beq \label{overlineG4_4d} 
\overline G_4 = \bigg[
d\gamma_\varphi \wedge V_2^\varphi
+ d\gamma_\Sigma \wedge V_2^\Sigma
\bigg] \wedge \frac{\widehat D \psi}{2\pi}
- (\chi \, \gamma_\varphi + 2 \, \gamma_\Sigma) \, 
V_2^\varphi \wedge V_2^\Sigma \ ,
\eeq
where 
$\gamma_\varphi$, $\gamma_\Sigma$ are functions of $\mu$ only and
$V^\varphi_2$, $V^\Sigma_2$ are proportional to the volume forms of $S^2_\varphi$ and the Riemann surface, respectively, normalized
according to
\beq \label{Vnormalization}
\int_{S^2_\varphi} V_2^{\varphi} = 1 \ , \qquad
\int_{\Sigma_g} V^\Sigma_2 = 1 \ .
\eeq
  The flux parameters are given as
\begin{equation} \label{gamma_eqs}
\gamma_\varphi^\mathrm N - \gamma_\varphi^\mathrm S = N \ , \quad 
\gamma_\Sigma^\mathrm N - \gamma_\Sigma^\mathrm S   =   N_\Sigma \ , \quad 
\chi \, \gamma_\varphi^\mathrm N + 2 \, \gamma_\Sigma^\mathrm N
=-N_\mathrm N \ , \quad
\chi \, \gamma_\varphi^\mathrm S + 2 \, \gamma_\Sigma^\mathrm S
=-N_\mathrm S  \ ,
\end{equation} where the $\gamma$ parameters are defined as $\gamma_{\bullet}^{\rm N,S} \equiv \gamma_{\bullet}(\mu_{\rm N,S})$.  The $\gamma$ relations imply\footnote{The constraint on the flux parameters can be understood as the condition that must be satisfied by the first Chern class of the $S^1_\psi$ bundle over the rest of the space:
\begin{equation}
\int_{\mathcal{C}_\mathrm N^4} \overline {G}_4 - \int_{\mathcal{C}_\mathrm S^4} \overline {G}_4 = \int_{M_6} \overline {G}_4\, d\frac{\widehat{D} \psi}{2\pi}.  
\end{equation} The minus sign is due the fact that the two cycles have opposite orientations in $M_6$.  }
\begin{equation}
2N_\Sigma + \chi N = - (N_\mathrm N - N_\mathrm S) \ .  
\end{equation}  The class $G_4$ is labeled by three flux parameters that we denote as $(N,N_\mathrm N, N_\mathrm S)$.  The flux parameter $N$ can be associated with the four-cycle $\mathcal{C}_\mathrm C^4$.

The flux parameters can be equivalently regarded as the coefficients of
the expansion of $\overline G_4$ onto coholomogy classes of $M_6$,
\beq
\overline G_4 = N \, \cV_4^\mathrm C + N_\mathrm N \, \cV_4^\mathrm N
+ N_\mathrm S \, \cV_4^\mathrm S + \text{(exact terms)} \ .
\eeq
The 4-forms $\cV_4^\alpha$, $\alpha = {\rm N,S,C}$ are closed but not exact,
and define a basis of cohomology classes in $M_6$.
We can parametrize them uniformly by writing
\beq \label{four_forms}
\cV_4^\alpha =
d\bigg[ U_\varphi^\alpha \, V_2^\varphi \wedge \frac{\widehat D\psi}{2\pi}
+ U_\Sigma^\alpha \, V_2^\Sigma \wedge \frac{\widehat D\psi}{2\pi}
\bigg]  + C^\alpha \, V_2^\varphi \wedge V_2^\Sigma \ .
\eeq
In the above expression, $U_\varphi^\alpha$, $U_\Sigma^\alpha$
are functions of $\mu$, while $C^\alpha$ is constant.
The parametrization \eqref{four_forms}
is subject to a 2-parameter redundancy,
related to shifts of $U_\varphi^\alpha$, $U_\Sigma^\alpha$ by constants.
A way to fix this redundancy is to demand
\beq
(U_\varphi^\alpha)^\mathrm N +(U_\varphi^\alpha)^\mathrm S = 0 \ , \qquad
  (U_\Sigma^\alpha)^\mathrm N +(U_\Sigma^\alpha)^\mathrm S = 0  \ .
\eeq
We want the 4-forms $\cV_4^\alpha$ to be dual to the 4-cycles
$\cC^4_\alpha$ defined above,
\beq
\int_{\cC^4_\alpha} \cV_4^\beta = \delta^\beta_\alpha \ .
\eeq
This condition determines the constants $C^\alpha$
and the quantities $(U_\varphi^\alpha)^\mathrm N$, 
$(U_\Sigma^\alpha)^\mathrm N$, according to the following table,
\beq \label{fourform_table}
\begin{array}{l  || c | c | ccc}
  & \phantom{l} (U_\varphi^\alpha)^\mathrm N \phantom{l}  & \phantom{l} (U_\Sigma^\alpha)^\mathrm N  \phantom{l} & \phantom{l} C^\alpha \phantom{l}   \\[1mm]  \hline
\alpha = \mathrm C \phantom{l}& \frac 12 & - \frac \chi 4 & 0   \\[1mm]
\alpha = \mathrm N \phantom{l}& 0 & - \frac 14 & \frac 12   \\[1mm]
\alpha = \mathrm S \phantom{l}& 0 &  \frac 14  & \frac 12    
\end{array}
\eeq

Another way to interpret the fluxes $(N_\mathrm N,N_\mathrm S)$ is to consider the fate of the fluctuations of the $C_3$ potential from the $SU(2)_\mathrm N\times SU(2)_\mathrm S$ symmetry before the compactification on the Riemann surface.   The curvatures associated to $SU(2)_\mathrm N\times SU(2)_\mathrm S$
are   $(F_{6d}^\mathrm N,F_{6d}^\mathrm S)$. 
We use the notation 
$(n_{6d}^\mathrm N,n_{6d}^\mathrm S)$
for the corresponding  Chern roots.
Since these forms have legs on the worldvolume directions of the M5-branes, $W_6$, when we compactify on the Riemann surface, they decompose as
\cite{Bah:2017gph}
\begin{equation} \label{fluxes_6d_to_4d}
n^\mathrm N_{6d}= N_\mathrm N \,    V_2^\Sigma     + 
N \, \frac{F^\mathrm N }{2\pi}, \qquad 
n^\mathrm S_{6d} = N_\mathrm S  \,    V_2^\Sigma  
+ N \,  \frac{F^\mathrm S }{2\pi} \ .  
\end{equation}
The quantities $(F^\mathrm N, F^\mathrm S)$ are 4d external
connections. We have introduced a factor $N$ in such a way that
both terms on the RHSs of \eqref{fluxes_6d_to_4d} scale linearly with the flux parameters $N$, $N^\mathrm N$, $N^\mathrm S$.
In the reduction, this decomposition implies the flux terms in $\overline {G}_4$ as 
\begin{equation}
\overline {G}_4 = N_\mathrm N  \,    V_2^\Sigma  \wedge \omega_\mathrm N +
 N_\mathrm S  \,    V_2^\Sigma  \wedge \omega_\mathrm S + \dots
\end{equation}
In field theory, the flux correspond to twisting the Cartan elements of the six-dimensional $SU(2)_\mathrm N \times SU(2)_\mathrm S$ symmetry over the Riemann surface.  

There are three harmonic two-forms associated to the four-cycles by Poincar\'{e} duality, we denote them as $(\omega_\mathrm N, \omega_\mathrm S, \omega_\mathrm C)$.  Indeed, the first two are just the resolution two cycles of $M_4$ which are preserved in $M_6$.  The flux quantization conditions can be equivalently written as
\begin{equation} \label{poincare_duality}
\int_{M_6} \overline {G}_4 \wedge \omega_\mathrm N = N_\mathrm N, \qquad \int_{M_6} \overline {G}_4 \wedge \omega_\mathrm S = N_\mathrm S, \qquad \int_{M_6} \overline {G}_4 \wedge \omega_\mathrm C = N.  
\end{equation} These formulas will be useful in the computation of the anomalies for the four-dimensional theories of interest.

In the reduction of M-theory on $M_6$, there are a class of fluctuations we can add for the $C_3$ potential.  For each of the harmonic two-forms, we can add a gauge field in the external spacetime given as $(a^\mathrm N, a^\mathrm S, a^\mathrm C)$ with field strength $(f^\mathrm N, f^\mathrm S, f^\mathrm C)$.  
We are using lowercase letters to emphasize that these are
gauge fields in 7d supergravity.
Naively, each one of these fields should lead to a $U(1)$ gauge symmetry on the external spacetime which then induces a $U(1)$ flavor symmetry on the worldvolume theory on the branes.  However, the $C_3$ potential can also have a three-form fluctuation, $c_3$, on the external spacetime with field strength, $g_4 = dc_3$.  These terms can be collected as
\begin{equation}
C_3 = a^\mathrm N \wedge \omega_\mathrm N + a^\mathrm S \wedge \omega_\mathrm S + a^\mathrm C \wedge \omega_\mathrm C + c_3+ \dots  
\end{equation} When we reduce the effective action of M-theory, the effective action of the seven-dimensional theory will have terms
\begin{equation} \label{7d_action}
S = \int  b_1 g_4 \wedge \star g_4 + b_2 N_\alpha a^\alpha \wedge g_4 + \dots
\end{equation} where the $b$'s are numbers.  We observe that $g_4$ couples to a linear combination of the gauge field.  This coupling implies that $g_4$ can be dualized to a St\"uckelberg field which is eaten by the gauge field $N_\alpha a^{\alpha}$.  Here and in what follows, the index $\alpha$ enumerates harmonic 2-forms
and takes the values $\alpha = \mathrm N, \mathrm S, \mathrm C$.
We refer the reader to \cite{Bah:2019rgq} for a more detailed
discussion of this St\"uckelberg mechanism.

\paragraph{Symmetries of the system}  The low-energy quantum field preserves a $U(1)_\mathcal{R}$ R-symmetry and the $SU(2)_\varphi$ flavor symmetries corresponding to the isometries of $S^1_\psi$ and $S^2_\varphi$.  Naively there is an $U(1)_\mathrm N \times U(1)_\mathrm  S \times U(1)_\mathrm  C$ corresponding to the fluctuations of $C_3$ along the harmonic two-forms $(\omega_\mathrm  N, \omega_\mathrm  S, \omega_\mathrm  C)$.  One linear combination is broken and only a $U(1)^2$ is preserved, which we denote as $U(1)_\mathrm  N \times U(1)_\mathrm  S$.  In the special cases when $g=0$, the Riemann surface is a two-sphere, in the low-energy limit it can admit an $SU(2)$ isometry group.  The flavor symmetry will further enhance by an $SU(2)$ flavor symmetry.  

Now we are in a position to construct the gauge invariant and globally defined boundary condition for the $G_4$ flux in presences of curved branes.  We have the action of the symmetries in $\overline {G}_4$ by adding suitable connection forms and curvature terms to make it closed:
\begin{align} \label{final_E4}
\overline {G}_4 \, \to \, E_4 &=   \big( 
d\gamma_\varphi \wedge e_2^\varphi
+ d\gamma_\Sigma \wedge e_2^\Sigma
\big) \wedge  \, 
\frac{\widetilde D \psi}{2\pi} 
+ \big( \gamma_\varphi \, e_2^\varphi
+ \gamma_\Sigma \, e_2^\Sigma \big) \wedge \bigg( 
-2 \, e_2^\varphi - \chi \, e_2^\Sigma + 2 \, \frac{F^\psi}{2\pi}
\bigg)
 \nn \\
&+N \,  \frac{F^\mathrm  N}{2\pi} \wedge  \widetilde \omega_\mathrm  N 
+  N \, \frac{F^\mathrm  S}{2\pi} \wedge \widetilde \omega_\mathrm  S
 + N \,  \frac{F^\mathrm  C}{2\pi} \wedge  \widetilde \omega_\mathrm  C \ , \qquad \mbox{with} \quad  F^\alpha N_\alpha =0 \ .
\end{align}
The last condition imposes the fact that one of gauge fields in the fluctuations of the $C_3$ potential is massive.  In the expression for $E_4$ we have gauged the isometry group as
\begin{align} \label{tildeDpsi}
\frac{\widetilde{D} \psi}{2\pi} = \frac{d \psi}{2\pi}  - 2 \,   \cA^\varphi  - \chi \,  \cA^{\Sigma}  + 2 \,  
 \frac{ A^\psi }{2\pi} \  , \qquad   d A^\psi = F^\psi ,  \qquad d \cA^\varphi = e_2^\varphi, \qquad
d \cA^\Sigma =   e^\Sigma_2 \ .  
\end{align}
The quantity $A_\psi$ is the 4d connection for   $U(1)_\psi$.
The two-forms, $e^\varphi_2$ and $e_2^\Sigma$ are the closed and gauge invariant volume forms of $S^2_\psi$ and $\Sigma_g$ respectively. 
The expression for $e_2^\varphi$~is
\beq \label{e2_def}
e_2^\varphi = \frac{1}{8\pi} \, \epsilon_{abc} \, (Dy^a \wedge Dy^b \, y^c - F^{ab} \, y^c)
\ , \qquad Dy^a = dy^a - A^{ab} \, y_b \ .
\eeq
The indices $a,b,c = 1,2,3$ are vector indices of $SO(3)_\varphi$,
raised and lowered with $\delta_{ab}$. The three quantities $y^a$
are constrained coordinates on $S^2_\varphi$, with $y^a \, y_a = 1$.\footnote{More explicitly,
\beq
y^1 = \sin \theta \, \cos \varphi \ , \qquad
y^2 = \sin \theta \, \sin \varphi \ , \qquad
y^3 = \cos \theta \ .
\eeq
}
The 1-forms $A^{ab}$ are the components of the external $SO(3)_\varphi$ connection,
and $F^{ab}$ are the components of the field strength.
When $g>1$, $e_2^\Sigma$ is simply the volume form on the Riemann surface,
$e_2^\Sigma = V_2^\Sigma$.
When $g = 0$ and the surface is a sphere, $e_2^\Sigma$
is given similarly as \eqref{e2_def}.
We need the integrals
\begin{equation} \label{integral_identities}
\int_{\Sigma_g} e_2^\Sigma = 1 \  , \qquad \int_{S^2} e_2^{2s+2} = 0 \ , \quad \int_{S^2} e_2^{2s+1} = 2^{-2s} \left[p_1(SO(3)) \right]^s, \quad s=0,1,2  , \dots
\end{equation} where $p_1(SO(3)_\varphi)$ is the first Pontryagin class of the $SO(3)_\varphi$ bundle over $W_4$.

The tilde over the harmonic 2-forms $\omega_{\mathrm N, \mathrm S, \mathrm C}$ 
in \eqref{final_E4}
signals the fact that we have gauged the isometry group
and we have restored closure, as explained below.

\paragraph{Harmonic 2-forms}  Before 
gauging the isometry group, the harmonic
2-forms $\omega_\alpha$, $\alpha  = \mathrm N, \mathrm S, \mathrm C$, can be uniformly 
parametrized as
\beq
\omega_\alpha = d \bigg[ H_\alpha(\mu) \, \frac{\widehat D\psi}{2\pi} \bigg]
+ t^\varphi_\alpha  \, V_2^\varphi 
+ t^\Sigma_\alpha \, V^\Sigma_2 \ ,
\eeq
where $t_\alpha^\varphi$, $t_\alpha^\Sigma$ are suitable constants
and $H_\alpha$ is a suitable function of $\mu$.
We promote $\omega_\alpha$ to $\widetilde \omega_\alpha$
by writing
\beq
\widetilde \omega_\alpha = d \bigg[ H_\alpha(\mu) \, \frac{\widetilde  D\psi}{2\pi} \bigg]
+ t^\varphi_\alpha  \, e_2^\varphi 
+ t^\Sigma_\alpha \, e^\Sigma_2 \ .
\eeq
This object is indeed manifestly closed and gauge-invariant.

Our parametrization of $\omega_\alpha$ is subject to a 1-parameter
redundancy related to shifts of $H_\alpha$ by a constant.
We fix this redundancy by demanding
\beq
H_\alpha^\mathrm N + H_\alpha^\mathrm S = 0 \ .
\eeq
Moreover, we want to basis $\omega_\alpha$ to be dual
to the basis $\cV^\alpha_4$ of 4-forms defined in \eqref{four_forms},
in the sense that
\beq
\int_{M_6} \cV_4^\alpha \wedge \omega_\beta = \delta^\alpha_\beta\  .
\eeq
This is achieved by fixing the quantities
$H_\alpha^\mathrm N$, $t_\alpha^\varphi$, $t_\alpha^\Sigma$
according to 
\beq \label{H_table}
\begin{array}{l  || c | c | ccc}
  & \phantom{l} H_\alpha^{\rm N} \phantom{l}  &
    \phantom{l} t_{\alpha}^\varphi \phantom{l} &
  \phantom{l}  t_{\alpha }^\Sigma \phantom{l} \\[1mm]  \hline
\alpha = \mathrm C \phantom{l}& 0 &   0 & 1 \\[1mm]
\alpha = \mathrm N \phantom{l}& \frac 12  & -1 & -\frac \chi 2  \\[1mm]
\alpha = \mathrm S \phantom{l}& \frac 12   & 1 & \frac \chi 2 
\end{array}
\eeq
This table summarize all information about $\omega_\alpha$
that is needed for the computation of  anomaly inflow in the next section.

\section{Anomalies for the low energy QFT}  \label{4d_anomalies}


Now we are in a position to compute anomalies for M5-branes probing the $k=2$ orbifold singularity.  The construction of the boundary data above will allow for an explicit computation for the anomaly polynomial for the field theories that describe the low energy dynamics of the branes \cite{Bah:2018gwc,Bah:2019jts,Bah:2019rgq}.  These are captured by a 12-form M-theory anomaly polynomial given as
\begin{equation} \label{I12_and_X8}
\mathcal{I}_{12} = - \frac{1}{6} E_4^3 - E_4 \wedge X_8, \qquad X_8 = \frac{1}{192} \left[p_1(TM_{11})^2  - 4\,  p_2(TM_{11}) \right]  \ ,
\end{equation} where $p_1(TM_{11})$, and $p_2(TM_{11})$ are the first and second Pontryagin classes of the tangent bundle of the eleven-dimensional M-theory spacetime, $TM_{11}$.  The four-form $E_4$ is precisely the gauge invariant boundary and globally defined boundary condition for the $G_4$ flux.  The anomaly six-form for the four-dimensional theories discussed above is given as
\begin{equation} \label{I12_integral}
I_6^{\rm inflow} = \int_{M_6} \mathcal{I}_{12}   \ ,
\end{equation} where we use the corresponding $E_4$ given in \eqref{final_E4}.  

%


The task at hand is the computation of the 8-form $X_8$
for the geometry \eqref{M6met} and of the integrals
$\int_{M_6} E_4 \, X_8$, $\int_{M_6} E_4^3$
with $E_4$ as in \eqref{final_E4}.
The full derivation is reported in detail in \cite{Bah:2019rgq}.
Here we point out some salient features of the analysis.

The 8-form $X_8$ is constructed with the first and second
Pontryagin classes of the 11d tangent bundle $TM_{11}$,
see \eqref{I12_and_X8}.
For the class of 4d theories under examination,
these classes can be computed using the following splitting
of $TM_{11}$,
\beq \label{bundle_split}
TM_{11} \rightarrow TW_4 \oplus T\Sigma_{g} \oplus 
TS^2_\varphi \oplus TS^2_{\mu\psi}  \  .
\eeq
The above expression is motivated recalling
that the space $M_6$ is a fibration of
$M_4$ (the resolved orbifold $S^4/\mathbb Z_2$)
over $\Sigma_g$, and that $M_4$
is a fibration of the 2-sphere $S^2_{\mu \psi}$ spanned by
$\mu$, $\psi$ over the 2-sphere $S^2_\varphi$.
The gauging of the isometry $SO(3)_\varphi$
shifts the Chern root of $TS^2_\varphi$
with a contribution with legs along $W_4$.
By a similar token, the Chern root of 
$TS^2_{\mu \psi}$ is shifted by the fact that $S^1_\psi$
is non-trivially fibered over $S^2_\varphi$, $\Sigma_g$,
and $W_4$, as can be inferred from the expression
of $\widetilde D\psi$ in \eqref{tildeDpsi}.
The split \eqref{bundle_split} implies 
\begin{align} \label{ps_eq}
p_1(TM_{11}) & = p_1(TW_4) + p_1(SO(3)_\varphi)
+ \bigg[ d  \frac{\widetilde D\psi}{2\pi} \bigg]^2 \  , \nn \\
p_2(TM_{11}) & = \big[  p_1(TW_4) + p_1(SO(3)_\varphi)
\big] \, \bigg[ d  \frac{\widetilde D\psi}{2\pi} \bigg]^2 \ .
\end{align}
We noticed that 
  terms with more than six external legs,
such as $p_1(TW_4) \,p_1(SO(3)_\varphi)$,
can be dropped, because they cannot
contribute to the the inflow anomaly
polynomial.
We notice that, if the Riemann surface is a sphere,
we can keep track of its $SO(3)_\Sigma$ isometry.
To this end, we simply have to replace
$ p_1(SO(3)_\varphi)$ with
$ p_1(SO(3)_\varphi) +  p_1(SO(3)_\Sigma)$
in \eqref{ps_eq}.


As pointed out earlier, the background curvatures
for the $U(1)$ symmetries associated to harmonic 2-forms
are subject to the constraint $F^\alpha N_\alpha = 0$.
This is due to the  
 argument given around \eqref{7d_action} for the 
emergence of a massive vector in 7d supergravity.
This argument, however,
 is
valid under the technical assumption 
\beq \label{condition}
\int_{M_6} \overline G_4 \wedge \omega_\psi = 0 \ .
\eeq
In the previous expression $\overline G_4$ is as in \eqref{overlineG4_4d}
and the 2-form $\omega_\psi$ is the coefficient of the linear term in $F^\psi$
inside $E_4$,
\beq
E_4 = F^\psi \wedge \omega_\psi + \dots \ , \qquad
\omega_\psi = 2 \, (2 \pi)^{-1} \, (\gamma_\varphi \, e_2^\varphi
+ \gamma_\Sigma \, e_2^\Sigma) \ .
\eeq
The requirement \eqref{condition}
ensures that the linear combination
of 7d $U(1)$ vectors  that gets massive via
St\"uckelberg mechanism
is built exclusively with the vectors
associated to harmonic 2-forms,
without any mixing with the vector $A^\psi$
associated to the isometry $U(1)_\psi$.
The interested reader can find a more
detailed discussion of this point in 
 \cite{Bah:2019rgq}.
If we combine \eqref{condition} with the relations
\eqref{gamma_eqs}, we can express the four quantities
$\gamma_\varphi^\mathrm N$, $\gamma_\varphi^\mathrm S$,
$\gamma_\Sigma^\mathrm N$, $\gamma_\Sigma^\mathrm S$
in terms of the three flux quanta
$(N, N_\mathrm N, N_\mathrm S)$,
\begin{align} \label{gamma_endpoints}
\gamma_\varphi^{\rm N,S} &= - \frac{N \, (N_\mathrm N + N_\mathrm S)}{ 2 \, (2 \chi N  + N_\mathrm N - N_\mathrm S )} \pm \frac 12 \, N \ , \\
\gamma_\Sigma^{\rm N,S} &=
- \frac{(N_\mathrm N + N_\mathrm S) (\chi N + N_\mathrm N - N_\mathrm S)}{4 \, (
2 \chi N + N_\mathrm N - N_\mathrm S
)} 
\mp \frac 14 (\chi N  + N_\mathrm N - N_\mathrm S)  \ .
\end{align}

It is worth noticing that the values of the integrals $\int_{M_6} E_4 \, X_8$,
$\int_{M_6} E_4^3$ are insensitive to the specific profile of the functions
$\gamma_\varphi$, $\gamma_\Sigma$, $H_\alpha$ entering $E_4$, but only depend on the values
that these functions attain at the endpoints of the $\mu$ interval.
For $\gamma_\varphi$ and $\gamma_\Sigma$
these values are given in \eqref{gamma_endpoints},
while for $H_\alpha$ they are collected in \eqref{H_table}.
We also notice that integration over $S^2_\varphi$ of powers of $e_2^\varphi$
 is conveniently
preformed making use of \eqref{integral_identities},
and similarly for integration along the Riemann surface.

After these preliminary remarks, we can give the full
expression for the inflow anomaly polynomial
$I_6^{\rm inflow}$ computed via \eqref{I12_integral}.
We   solve the constraint $F^\alpha N_\alpha = 0$
by expressing $F^\mathrm C$ in \eqref{final_E4} in terms of $F^{\rm N,S}$,
$F^\mathrm C = - \frac 1N \, (N_\mathrm N \, F^\mathrm N
+ N_\mathrm S \, F^\mathrm S) $.
We introduce the notation
\beq
c_1^\psi \equiv c_1(U(1)_\psi) = \frac{F^\psi}{2\pi} \ , \qquad
c_1^{\rm N,S} \equiv c_1(U(1)_{\rm N,S} ) = \frac{F^{\rm N,S}}{2\pi} \ , 
\eeq
and we write $I_6^{\rm inflow}$ in terms of
the quantities
\beq \label{other_quantities}
N_\Sigma = - \frac 12 ( N \, \chi + N_\mathrm N - N_\mathrm S  ) \ , \qquad
M = \frac 12 (N_\mathrm N + N_\mathrm S) \ .
\eeq
We find
\begin{align} \label{total_inflow}
- I_6^{\rm inflow} & = \bigg[
\frac{N^2 N_\Sigma (\chi N  - 2 N_\Sigma + 2M)(\chi N - 2 N_\Sigma - 2M)}{(\chi N - 2 N_\Sigma)^2}
+ \frac{\chi N}{2}
\bigg] \, (c_1^\psi)^2 \,
(c_1^\mathrm N + c_1^\mathrm S)
  \\
& - \frac{N^2 (\chi N - 2 N_\Sigma + 2 M)(\chi N + 2N_\Sigma - 2 M)}{2(\chi N - 2 N_\Sigma)} \, c_1^\psi \, (c_1^\mathrm N)^2
\nn \\
& - \frac{N^2 (\chi N + 2 N_\Sigma + 2 M) (\chi N - 2 N_\Sigma - 2 M)}{2(\chi N - 2 N_\Sigma)} \, c_1^\psi \, (c_1^\mathrm S)^2
\nn \\
& - \frac{N^2 (\chi^2 N^2 - 4 N_\Sigma^2 - 4 M^2)}{\chi N - 2 N_\Sigma} \, 
c_1^\psi \, c_1^\mathrm N \, c_1^\mathrm S
\nn \\
& + \frac 16 N^2 (\chi N - 6 N_\Sigma + 6M) \, (c_1^\mathrm N)^3
+ \frac 16 N^2 (\chi N - 6 N_\Sigma - 6M) \, (c_1^\mathrm S)^3
\nn \\
& + \frac 12 N^2 (\chi N + 2 N_\Sigma + 2 M) \, (c_1^\mathrm N)^2 \, c_1^\mathrm S
+ \frac 12 N^2 (\chi N + 2 N_\Sigma - 2 M) \, (c_1^\mathrm S)^2 \, c_1^\mathrm N
\nn \\
& + \frac{\chi N + 2 N_\Sigma}{24} \, c_1^\psi \, p_1(TW_4)
- \frac{\chi N}{24} (c_1^\mathrm N + c_1^\mathrm S) \, p_1(TW_4)
- \frac{\chi N + 2 N_\Sigma}{6} \, (c_1^\psi)^3
\nn \\
& + \bigg[
- \frac{N^2 (\chi^2 N^2 - 12 N_\Sigma^2 + 4 \chi N N_\Sigma + 12 M^2)}{24 (\chi N - 2 N_\Sigma)}
- \frac{\chi N}{12}
\bigg] \, c_1^\psi \, p_1(SO(3)_\varphi)
\nn \\
& + \frac 18 N^2 (\chi N + 2 N_\Sigma - 2 M) \, c_1^\mathrm N \, p_1(SO(3)_\varphi)
+ \frac 18 N^2 (\chi N + 2 N_\Sigma + 2 M) \, c_1^\mathrm S \, p_1(SO(3)_\varphi) 
\nn \\
& + \bigg[
- \frac{N_\Sigma^2 (3N^2 - N_\Sigma^2 - 2 N N_\Sigma - 3 M^2)}{12(N - N_\Sigma)}
- \frac{N_\Sigma}{6}
\bigg] \, c_1^\psi \, p_1(SO(3)_\Sigma)
\nn \\
& + \frac 14 N_\Sigma (N + M) (N + N_\Sigma - M) \, c_1^\mathrm N \, p_1(SO(3)_\Sigma)
\nn \\
& + \frac 14 N_\Sigma (N - M)(N + N_\Sigma + M) \, c_1^\mathrm S \, p_1(SO(3)_\Sigma)
\nn \ .
\end{align}
In the last three lines we have collected the terms
related to the symmetry $SO(3)_\Sigma$,
which is only present if the Riemann surface is a sphere.

\section{Holographic solutions}   \label{holo_section}

In the previous sections we have adopted a UV
point of view:  we fixed the supersymmetric M-theory
background \eqref{11d_background}
and we inserted a stack of M5-branes 
extended along $W_4 \times \Sigma_g$
and sitting at the origin of $\mathbb C^2/\mathbb Z_2 \times \mathbb R$, specifying also the appropriate
background fluxes along the Riemann surface.
In this section we argue that, in the large-$N$ limit,
this class of UV setups corresponds in the IR 
to a well-known class of $AdS_5$ solutions
in M-theory, first described in GMSW \cite{Gauntlett:2004zh}.

In the vicinity of the M5-branes,
the UV picture of 11d spacetime 
is described in \eqref{near_the_branes}.
Our expectation for the near-horizon IR picture, based on \cite{Gauntlett:2006ux},
is that the overall radial direction $\mathbb R^+$
combines with $W_4$ to yield
an $AdS_5$ factor, leaving the geometry $M_6$
as internal space. Taking into account backreaction
effects, the Ansatz for the 11d metric in the near-horizon limit has the form
\beq
ds^2(M_{11}) = e^{2\lambda} \, \big[ ds^2(AdS_5)
+ ds^2(M_6) \big] \ ,
\eeq
with $ds^2(M_6)$ of the form \eqref{M6met},
and $\lambda$ a warp factor depending on $M_6$.
Let us stress that all metric functions
in the ansatz \eqref{M6met} for the metric on $M_6$
depend on the interval coordinate $\mu$ only.
It is natural to also demand that
the warp factor $\lambda$ be a function of $\mu$ only.

In \cite{Gauntlett:2004zh} a class of solutions is described,
in which $M_6$ and $\lambda$ have exactly 
the properties described in the previous paragraph.
More precisely, the fully backreacted geometry of $M_6$ is
\beq \label{IRgeometry}
ds^2(M_6)  = e^{-6\lambda} \Big[ F_{\varphi} \, d  s^2(S^2_\varphi)
+ F_\Sigma \, d  s^2(\Sigma_g) \Big]
+ \frac{e^{-6\lambda}}{\cos^2\zeta} \,  d\mu^2
+ \frac{\cos^2 \zeta}{9} \,  \widehat D\psi^2 \ ,
\eeq
where the warp factor $\lambda$ and the metric functions
$F_\varphi$, $F_\Sigma$, $\cos \zeta$ depend on $\mu$ only.
Their expressions are recorded in appendix \ref{sugra_app},
where we summarize some key features of the GMSW solutions. 
The $G_4$-flux configuration of the holographic solution
is given in \eqref{GMSW_review_flux}. As expected, it has exactly the same structure
as $\overline G_4$ in \eqref{overlineG4_4d}.

The fact that the topology of the internal space $M_6$
in the GMSW solutions matches exactly
with the topology of $M_6$ in our UV setup
is a strong hint that the GMSW solution provides
the gravity dual to the field theory setups
we discussed in section  \ref{4dsection}.
Furthermore, the GMSW solutions provide evidence
for the fact that the 4d construction yields a non-trivial
IR fixed point, at least at large $N$.

In the remainder of this section we perform two
quantitative checks of our proposed field theory interpretation
of the GMSW solutions. Before entering the details of the computation,
let us briefly discuss our stategy.

On the field theory side, 
the inflow anomaly polynomial \eqref{total_inflow} is expected to be exact in $N$,
but to contain both the anomalies of the interacting SCFT of interest
and of decoupled sectors. It is natural to assume that the decoupled sectors
do not contribute to the leading order $N^3$. As a result, from \eqref{total_inflow}
we can safely extract the anomaly polynomial
for the interacting SCFT at large $N$.
We then perform $a$-maximization \cite{Intriligator:2003jj} at large $N$ in order to identify
which linear combination of $U(1)_\psi$, $U(1)_{\rm N,S}$ is the
superconformal R-symmetry.
Once the latter is determined, its 't Hooft anomaly coefficients
give us the central charge $c$ and   the flavor central charge $B$
for the $SO(3)_\varphi$ symmetry originating from isometries of $S^2_\varphi$.
The quantities $c$ and $B$ can also be computed holographically
in the GMSW solutions. This supergravity computation is reported in
appendix \ref{sugra_app}. We find a perfect agreement with the
field theory results.

Let us discuss in greater detail the field theory derivation
of the quantities $c$, $B$.
The first step is simply to isolate the leading terms in \eqref{total_inflow}
at large $N$. For simplicity, in this section we do not keep
track
of the $SO(3)_\Sigma$ symmetry that is present in the case
in which the Riemann surface is a sphere.
We may then write
\begin{align} \label{large_N_poly}
I_6^\text{SCFT, large $N$}& =  
\frac{N^2 N_\Sigma (\chi N  - 2 N_\Sigma + 2M)(\chi N - 2 N_\Sigma - 2M)}{(\chi N - 2 N_\Sigma)^2}
\, (c_1^\psi)^2 \,
(c_1^\mathrm N + c_1^\mathrm S)
  \\
& - \frac{N^2 (\chi N - 2 N_\Sigma + 2 M)(\chi N + 2N_\Sigma - 2 M)}{2(\chi N - 2 N_\Sigma)} \, c_1^\psi \, (c_1^\mathrm N)^2
\nn \\
& - \frac{N^2 (\chi N + 2 N_\Sigma + 2 M) (\chi N - 2 N_\Sigma - 2 M)}{2(\chi N - 2 N_\Sigma)} \, c_1^\psi \, (c_1^\mathrm S)^2
\nn \\
& - \frac{N^2 (\chi^2 N^2 - 4 N_\Sigma^2 - 4 M^2)}{\chi N - 2 N_\Sigma} \, 
c_1^\psi \, c_1^\mathrm N \, c_1^\mathrm S
\nn \\
& + \frac 16 N^2 (\chi N - 6 N_\Sigma + 6M) \, (c_1^\mathrm N)^3
+ \frac 16 N^2 (\chi N - 6 N_\Sigma - 6M) \, (c_1^\mathrm S)^3
\nn \\
& + \frac 12 N^2 (\chi N + 2 N_\Sigma + 2 M) \, (c_1^\mathrm N)^2 \, c_1^\mathrm S
+ \frac 12 N^2 (\chi N + 2 N_\Sigma - 2 M) \, (c_1^\mathrm S)^2 \, c_1^\mathrm N
\nn \\
&  
- \frac{N^2 (\chi^2 N^2 - 12 N_\Sigma^2 + 4 \chi N N_\Sigma + 12 M^2)}{24 (\chi N - 2 N_\Sigma)}
 \, c_1^\psi \, p_1(SO(3)_\varphi)
\nn \\
& + \frac 18 N^2 (\chi N + 2 N_\Sigma - 2 M) \, c_1^\mathrm N \, p_1(SO(3)_\varphi)
+ \frac 18 N^2 (\chi N + 2 N_\Sigma + 2 M) \, c_1^\mathrm S \, p_1(SO(3)_\varphi) \ .
\nn
\end{align}
Next, we perform $a$-maximization.
The trial superconformal R-symmetry is a linear combination of $U(1)_\psi$
 with $U(1)_{\rm N,S}$, parametrized as
\beq \label{generator_shift}
R = T_\psi +   s^\mathrm N \, T_\mathrm N +   s^\mathrm S \, T_\mathrm S \ ,
\eeq
where $T_\psi$, $T_{\rm N,S}$ denote 
the   generators of $U(1)_\psi$, $U(1)_{\rm N,S}$,
and 
$  s^{\rm N,S}$ are parameters   to be fixed. 
At leading order at large $N$,
\beq
I_6^\text{SCFT} = \frac 16 \, {\rm tr} \, R^3 \, (c_1^R)^3 \ , \qquad
a = c = \frac{9}{32} \,  {\rm tr} \, R^3 \ , 
\qquad \qquad \text{(large $N$)}
\eeq
where $c_1^R$ is the first Chern class of the background curvature
for the superconformal R-symmetry.
At the level of the anomaly polynomial,
\eqref{generator_shift} is equivalent to 
the replacements
\beq \label{replacements}
c_1^\psi \rightarrow c_1^R \ , \qquad
c_1^{\rm N,S} \rightarrow   s^{\rm N,S} \, c_1^R \ .
\eeq
It follows that $a$-maximization at large $N$
can be carried out
by taking \eqref{large_N_poly}, performing the replacements
\eqref{replacements}, and maximizing the coefficient of $(F^R)^3$ with respect to the parameters
$  s^{\rm N,S}$.
The result of this computation is
most conveniently written in terms of the 
quantities $N_\Sigma$, $M$ defined in \eqref{other_quantities}.
The central charge reads
\begin{align} \label{largeN_c}
c & = 
\frac{9 N^2   N_{\Sigma }^2  
   \,      \Big[ \chi^2 N^2   + 2 \chi   N    N_\Sigma  + 4 \, N_\Sigma^2 - 3  M^2 \Big]^{3/2}    
   }{4 \left(3 M^2+2 \chi N    N_{\Sigma
   }\right){}^2}
   \nn \\
   & -\frac{9 N^2 N_{\Sigma }^2 \left(2 N_{\Sigma }+\chi N   \right)
   \left(2 \chi^2 N^2  +2 \chi N    N_{\Sigma }+8 N_{\Sigma }^2  -9 M^2 \right)}{8
   \left(3 M^2+2 \chi N    N_{\Sigma }\right){}^2} \ .
\end{align}
For completeness, let us also record the values of the 
parameters $s^{\rm N,S}$,
\begin{align} \label{s_parameters}
s^{\rm N,S} & = 
\frac{  2 \chi^2 N^2 N_\Sigma - 8 N_\Sigma^3  \pm M \, ( \chi^2 N^2 - 4 \chi NN_\Sigma - 4 N_\Sigma^2 )  \mp 6 M^3 }{ 2 \, (\chi N - 2 N_\Sigma) \, ( 3M^2 + 2 \chi N N_\Sigma ) }
\nn \\
& - \frac{2 N_\Sigma \mp M}{2 \, ( 3M^2 + 2 \chi N N_\Sigma)} \, \sqrt{  \chi^2 N^2   + 2 \chi   N    N_\Sigma  + 4 \, N_\Sigma^2 - 3  M^2   }  \ .
\end{align}

Let us now discuss the flavor central charge $B$
for the $SO(3)_\varphi$ symmetry.
The quantity $B$ appears in the 2-point function
of two $SO(3)_\varphi$ symmetry currents.
For its normalization, we follow the conventions of \cite{Freedman:1998tz}.
The superconformal algebra relates $B$ to the 't Hooft anomaly
between the superconformal R-symmetry and $SO(3)_\varphi$ \cite{Anselmi:1996dd,Anselmi:1997rd}.
Let us define the 't Hooft anomaly coefficient
$\cA_{SO(3)_\varphi}$ by
\beq
I_6^{\rm SCFT} = \cA_{SO(3)_\varphi} \, c_1^R \, p_1(SO(3)_\varphi) + \dots
\eeq
We then have
\beq
B = 4 \, \cA_{SO(3)_\varphi} \ .
\eeq
The quantity $\cA_{SO(3)_\varphi}$ is extracted from 
\eqref{large_N_poly} by performing the replacements
\eqref{replacements} and using \eqref{s_parameters}.
The result for $B$ then reads
\begin{align} \label{field_theory_B}
B  & = 
\frac{N^2 \left( 4 \chi^2 N^2   N_{\Sigma }+12 \chi N
      N_{\Sigma }^2+24 N_{\Sigma }^3
   -30 M^2 N_{\Sigma }-9 \chi M^2 N  \right)}{6 \left(3 M^2+2\chi  N    N_{\Sigma
   }\right)}
   \nn \\
   & -\frac{N^2   \left( \chi N    N_{\Sigma }+2 N_{\Sigma
   }^2  -M^2 \right)}{  3 M^2+2 \chi N    N_{\Sigma } } \, 
      \sqrt{   \chi^2 N^2   + 2 \chi   N    N_\Sigma  + 4 \, N_\Sigma^2 - 3  M^2   }  \ .
\end{align}

For definiteness, the supergravity computation of appendix \ref{sugra_app}
is performed in the case in which the Riemann surface has genus $g \ge 2$ and
the flux parameter $M$ is set to zero.
The results for $c$ and $B$ are given in \eqref{sugra_c}, \eqref{sugra_B}
in terms of the quantity $\tilde r = - 2 N_\Sigma /(N \chi)$.
They agree perfectly with \eqref{largeN_c}, \eqref{field_theory_B},
respectively.

\subsection{Comments on the reduction of the 6d anomaly polynomial}

In this section we contrast the approach of section \ref{4d_anomalies}
with the direct reduction on the Riemann surface of the anomaly
polynomial of the parent 6d (1,0) theory.
More precisely, let us consider the inflow anomaly polynomial
$I_8^{\rm inflow}$ in \eqref{6d_inflow}, and let us integrate it on $\Sigma_g$.
To this end, it is useful to express $c_2^\psi$ and ${\rm tr} (F^{\rm N,S}_{6d})^2$
in terms of 6d Chern roots. Following \cite{Bah:2017gph}, we have
\beq
c_2^\psi = -  (n^\psi_{6d})^2 \ , \qquad
{\rm tr} \, \frac{(F^{\rm N,S}_{6d})^2}{(2\pi)^2} =  - 4 \, (n^{\rm N,S}_{6d})^2   \ .
\eeq
The 6d Chern roots split as
\beq
n^\psi_{6d} = c_1^\psi - \frac \chi 2 \, V_2^\Sigma \ , \qquad
n^{\rm N,S}_{6d} = N \, c_1^{\rm N,S} + N_{\rm N,S} \, V_2^\Sigma \ ,
\eeq
with $V_2^\Sigma$ normalized as in \eqref{Vnormalization}.
We are not twisting $SU(2)_\varphi$, whose connection is thus
purely external.
Upon integration on $\Sigma_g$, we   obtain
\begin{align} \label{inflow_integrated}
- \int_{\Sigma_g} I_8^{\rm inflow} & =
- \frac 13 \, \chi \, \left (N^3 - \frac 32 \, N \right) \, (c_1^\psi)^3
- 2 \, N^2 \,  (N_\mathrm N \, c_1^\mathrm N 
+ N_\mathrm S \, c_1^\mathrm S  ) \, (c_1^\psi)^2
\nn \\
& + \chi \, N^3 \, c_1^\psi \, \big[  (c_1^\mathrm N)^2 + (c_1^\mathrm S)^2 \big]
- 2 \, N^2 \, \big[ N_\mathrm N \, (c_1^\mathrm N)^3
+ N_\mathrm S \, (c_1^\mathrm S)^3  \big]
\nn \\
&+ 2 \, N^2 \, \big[
N_\mathrm N \, c_1^\mathrm N \, (c_1^\mathrm S)^2
+ N_\mathrm S \, c_1^\mathrm S \, (c_1^\mathrm N)^2
\big]
- \frac{1}{24} \, \chi \, N \, c_1^\psi \, p_1(TW_4) 
\nn \\
&  - \frac 13 \, \chi \, (N^3 - N) \, c_1^\psi \, c_2^\varphi
- 2 \, N^2 \, (N_\mathrm N \, c_1^\mathrm N 
+ N_\mathrm S \, c_1^\mathrm S  ) \, c_2^\varphi
\ .
\end{align}
If we perform $a$-maximization at large $N$
using \eqref{inflow_integrated} as an input, we get a 
central charge $c$  that does not agree with  
 \eqref{largeN_c}.
Working for simplicity in the case $M =0$, or equivalently 
$N^\mathrm N + N^\mathrm S = 0$,
we obtain the results
\beq
c = -  \frac{9 \, N \, ( 5 \chi^2 N^2 + 12 \chi N N_\Sigma + 12 N_\Sigma^2 )}{32 \chi}  \ ,
\qquad
s^{\rm N,S}  = \mp \frac{N_\Sigma}{\chi N} \mp \frac 12 \ , \qquad
(M=0)
\eeq
which have a different structure compared to
\eqref{largeN_c}, \eqref{s_parameters} at $M = 0$,
due to the absence of radicals.
We have verified numerically in a few examples that
the discrepancy between the correct central charge \eqref{largeN_c}
and the central charge obtained from \eqref{inflow_integrated}
persists for $M \neq 0$.
This test can be regarded as a basis-independent check
that \eqref{inflow_integrated} and \eqref{total_inflow} 
are inequivalent anomaly polynomials.

If the Riemann surface has genus $g \ge 2$,
we can consider the limit $M = 0$, $N_\Sigma = - \frac \chi 2 \, N$.
This is equivalent to setting $N^{\rm N,S} = 0$.
As a result, we are blowing down the resolution 2-cycles,
and the geometry re-develops orbifold singularities.
In this scenario, the reduction of the 6d anomaly polynomial
gives a large-$N$ central charge
that agrees with our 4d inflow anomaly polynomial \eqref{total_inflow}.
We detect, however, a mismatch in the 't Hooft anomaly coefficients
for $U(1)_{\rm N,S}$. We interpret this discrepancy as
being due to decoupled modes in the resolved phase,
which have to be re-included in the limit 
$N^{\rm N,S} \rightarrow 0$.


\section{Discussion} \label{discussion_sec}

In this work we have mainly focused on 
the 6d (1,0) theory living on a stack of $N$ M5-branes probing
a $\mathbb C^2/\mathbb Z_2$ singularity.
We expect, however, that many features of this setup
should persist for branes probing 
$\mathbb C^2/\mathbb Z_k$ for $k \ge 3$.
By resolving the orbifold singularities at the north and south poles
of $S^4/\mathbb Z_k$, the flavor symmetry $SU(k)_\mathrm N \times SU(k)_\mathrm S$ is broken to $[U(1)^{k-1}]_\mathrm N \times [U(1)^{k-1}]_\mathrm S$.
We can then compactify on a Riemann surface
with a non-trivial twist for this symmetry.
We  expect  the emergence of an accidental
$U(1)$ symmetry and the spontaneous breaking of a $U(1)$ generator
to occur in such setups.

In the case $k = 2$ the geometry of the resolution of $S^4/\mathbb Z_2$
is particularly simple. This facilitates the identification
of the gravity duals. Nonetheless, it would be interesting to
investigate the gravity duals also for $k \ge 3$. In this case,
the internal geometry $M_4$,
associated to the 6d QFT in its resolved phase, is expected to have a smaller
$U(1)_\psi \times U(1)_\varphi$ isometry, and a more complicated
topology. The identification of the dual $AdS_5$ solutions
would be particularly useful, since it would allow us to perform
large-$N$ supergravity tests similar to the ones 
considered in this work for $k=2$.  These solutions should be obtained from BPS system described in \cite{Bah:2015fwa}.

The examples studied in this paper show the
power of geometric methods in the study of strongly
coupled dynamics of 4d QFTs.  In particular,
by constructing the 4-form $E_4$ 
that governs anomaly inflow from the M-theory ambient
space, we are able to track directly
the emergence of accidental symmetries and 
spontaneous symmetry breaking. Our analysis
fits into a broader  geometrization program,
aimed at using geometric and string theoretic tools  define
and classify
non-trivial QFTs, and to uncover their
non-perturbative dynamics.

\section*{Acknowledgments}

We would like to thank Chris Beem, Ken Intriligator, Ruben Minasian, Emily Nardoni, Alessandro Tomasiello, Peter Weck for interesting conversations and correspondence. The work of IB and FB is supported in part by NSF grant PHY-1820784.  We gratefully acknowledge the Aspen Center for Physics, supported by NSF grant PHY-1607611, for hospitality during part of this work.  


\appendix


\section{Supergravity computations}  \label{sugra_app} 

\subsection{Review of the GMSW solutions}

In this appendix we review a class of  M-theory 
solutions with 4d $\cN = 1$ superconformal symmetry,  first described in GMSW \cite{Gauntlett:2004zh}.
The 11d metric reads
\begin{align} \label{GMSW_review_metric}
ds^2_{11} & =  e^{2\lambda} \, \Big[ ds^2(AdS_5) 
+  ds^2(M_6) \Big] \ ,  \nn  \\
ds^2(M_6) & = e^{-6\lambda} \Big[ F_{\varphi} \, d  s^2(S^2_\varphi)
+ F_\Sigma \, d  s^2(\Sigma_g) \Big]
+ \frac{e^{-6\lambda}}{\cos^2\zeta} d\mu^2
+ \frac{\cos^2 \zeta}{9} \widehat D\psi^2 \ , \nn \\
d\widehat D\psi &= (2\pi) \, ( -2 \, V_2^\varphi - \chi \, V_2^\Sigma ) \ , \qquad
\int_{S^2_\varphi} V_2^\varphi = 1 \ , \qquad
\int_{\Sigma_g} V_2^\Sigma =   1  \ .
\end{align}
We have set the $AdS_5$ radius to 1, so that
the Ricci scalar of $AdS_5$ is $R  = -20$.
The metric on the Riemann surface $\Sigma_g$ has curvature $k = \pm 1$,
with Ricci scalar $R = 2 \, k$.
Compared with \cite{Gauntlett:2004zh}, we have flipped the sign of $\psi$
and we have renamed $y$ into $\mu$.
All metric functions depend on $\mu$ only.
They are given by
\begin{align}
e^{6\lambda} &= 
\frac{2   \left(a_\varphi -     \mu^2\right) \left(a_\Sigma - k  \, 
   \mu^2\right)}{a_\Sigma   +k \, a_\varphi  +2  \, k  \, \mu  \,  \left(  \mu - 3 \, \gamma_0
    \right)}
    \ , \qquad 
F_\varphi   = \frac 13 (a_\varphi -   \mu^2)    \ , \qquad
F_\Sigma   = \frac 13 (a_\Sigma - k  \,  \mu^2) \ ,
\end{align}
where $a_\varphi$, $a_\Sigma$, $\gamma_0$
are constant parameters. The quantity $0 \le \cos \zeta \le 1$
is determined by
\beq
e^{3\lambda} \, \sin \zeta = 2 \, \mu \ .
\eeq
The $G_4$-flux configuration is given by
\begin{align} \label{GMSW_review_flux}
G_4 &=2\pi \, d \Big[  \big( \widetilde  \gamma_\varphi \, V_2^\varphi
+ \widetilde \gamma_\Sigma \, V_2^\Sigma \big) \wedge \widehat D \psi \Big]
\nn \\
& = 2\pi \, 
\Big[ d\widetilde \gamma_\varphi \wedge V_2^\varphi + d\widetilde \gamma_\Sigma \wedge   V_{2}^\Sigma  \Big] \wedge \widehat D\psi 
 - (2\pi)^2 \, (2 \, \widetilde \gamma_\Sigma + \chi  \,\widetilde  \gamma_\varphi +
2 \, \chi \,   \gamma_0) \, V_2^\varphi \wedge    V_2^\Sigma
 \ ,
\end{align}
with the functions $\widetilde \gamma_\varphi$, $\widetilde \gamma_\Sigma$  given as
\begin{align} \label{fi_expr}
\widetilde \gamma_\varphi & = 2 \, \frac{
2  \, k \, a_\Sigma \,     \mu - 6  \, k \, a_\Sigma  \, \gamma_0 
+a_\varphi \,  \mu +  \mu^3
}{
9  \, k  \,   \left(a_\Sigma  - k  \, \mu^2 \right)
}  
\ , \nn \\
\widetilde \gamma_\Sigma & =  \chi \, \frac{
2 \,  k \, a_\varphi \,    \mu -  6  \, k \, a_\varphi \,  \gamma_0 +a_2 \,  y
+ k  \, \mu^3
}{
9 \, k \,  \left(a_\varphi -   \mu^2\right)
} 
\ .
\end{align}
We have put a tilde on $\widetilde \gamma_\varphi$, $\widetilde \gamma_\Sigma$
to distinguish these functions, coming from the holographic solution,
from the functions $\gamma_\varphi$, $\gamma_\Sigma$
that enter the parametrization \eqref{final_E4} of $E_4$ in the main text.
In this appendix, 
we are adopting conventions in which the quantization of $G_4$ flux 
reads
\beq \label{G4_quantization_convention}
\int_{\cC_4} \frac{G_4}{(2\pi \ell_p)^3} \in \mathbb Z \ ,
\eeq
where $\cC_4$ is a 4-cycle and $\ell_p$ is the 11d Planck length.

Let us now focus on solutions with $\gamma_0 = 0$.
We verify that in this class of solutions the fluxes
$N^{\rm N,S}$ defined in \eqref{N_north_south_def} satisfy
\beq
N^\mathrm N + N^\mathrm S = 0   \ .
\eeq
The range of the coordinate $\mu$  is determined
from the zeros of $\cos \zeta$ and has the form
$[- \mu_{\rm N}, \mu_{\rm N}]$, with $\mu_{\rm N} >0$.
We find it useful to distinguish the cases in which
the genus $g$ of the Riemann surface is $g \ge 2$ contrasted to $g = 0$.
We find
\begin{align} \label{muN_expr}
g &= 0  \; : &
\mu_{\rm N}^2 &= - \frac 12 \, (a_\varphi + a_\Sigma) + \frac 12 \, \sqrt{
(a_\varphi + a_\Sigma)^2 + \frac 43 \, a_\varphi \, a_\Sigma
} \ , & 
&a_\varphi, a_\Sigma    >0   \ , \nn \\
g & \ge 2 \; : &
\mu_{\rm N}^2 &= \frac 12 \, (a_\Sigma - a_\varphi) - \frac 12 \, \sqrt{
(a_\Sigma - a_\varphi )^2 - \frac 43 \, a_\varphi \, a_\Sigma
} \ , &
&a_\Sigma  > 3 \, a_\varphi >0 \ .
\end{align}
Recall that the fluxes $N$ and $N_\Sigma$ are defined as
\beq
N = \int_{S^2_\varphi \times S^2_{\mu \psi}} \frac{G_4}{(2\pi \ell_p)^3} \ , \qquad
N_\Sigma = \int_{\Sigma_g \times S^2_{\mu \psi}} \frac{G_4}{(2\pi \ell_p)^3} \ .
\eeq
We can express the ratio $a_\Sigma/a_\varphi$ in terms of the ratio
$N_\Sigma/N$,
\begin{align} \label{ratioVSratio}
g &  = 0 \; : &
\frac{a_\Sigma}{a_\varphi} & = \frac{  2 \, r^2 - r + 2 + 2 \, (r-1) \, \sqrt{r^2 + r + 1}  }{3 \, r}  \ , & 
r &:= \frac{N_\Sigma}{N} \ , \nn \\
g &  \ge 2  \; : &
\frac{a_\Sigma}{a_\varphi} & = \frac{  2 \, \tilde r^2 + \tilde r + 2  + 2 \, \sqrt{  \tilde r^4 + \tilde r^3 + \tilde r + 1  } }{ 3 \, \tilde r} \ , & 
\tilde r & := - \frac{2 \, N_\Sigma}{\chi \, N} > 1  \ .
\end{align}
Moreover, we can express the ratio between the Planck length
and the $AdS_5$ scale (which was set to 1 in the line element) in terms of 
$a_\Sigma$, $a_\varphi$, $\mu_{\rm N}$, $N$,
\begin{align} \label{ellpcube}
g &  = 0 \; : &
\frac{\ell_p^3}{L_{AdS}^3} & = \frac{ 2 \, \mu_{\rm N} \, (2 \, a_\Sigma + a _\varphi + \mu_{\rm N}^2 ) }{9 \, \pi \, N \, (a_\Sigma - \mu_{\rm N}^2)}   \ , \nn \\
g &  \ge 2  \; : &
\frac{\ell_p^3}{L_{AdS}^3} & = \frac{2 \, \mu_{\rm N} \, (2 \, a_\Sigma - a_\varphi - \mu_{\rm N}^2 )}{ 9 \, \pi \, N \, (a_\Sigma + \mu_{\rm N}^2)}   \ .
\end{align}

\subsection{Effective action in five dimensions}
In order to compute holographically the central charge $c$
and the flavor central charge for the $SO(3)_\varphi$ isometry
of $S^2_\varphi$, we need to extract the coefficients of the Einstein-Hilbert
term in the 5d effective action, as well as the coefficient of the kinetic
terms for the $SO(3)_\varphi$ vectors.
To this end, we only need two terms in the 11d M-theory action,
\beq \label{Mtheory_action}
S_{11} = \frac{1}{2\kappa_{11}^2} \, \int_{M_{11}} \bigg[
R_{(11)} \, *_{11} 1 - \frac 12 \, G_4 \wedge *_{11} G_4 + \dots
\bigg] \ , \qquad
2 \kappa_{11}^2 = (2\pi)^8 \,\ell_p^9 \ .
\eeq
The dimensional reduction from 11d to 5d is performed
activating the external 5d metric and the gauge fields for $SO(3)_\varphi$.
The 11d line element then reads
\begin{align}  
ds^2_{11} & =  e^{2\lambda} \, \Big[ ds^2(M_5) 
+  ds^2(M_6) \Big] \ ,  \nn  \\
ds^2(M_6) & = e^{-6\lambda} \Big[ F_{\varphi} \, d  s^2(S^2_\varphi)^{\rm g}
+ F_\Sigma \, d  s^2(\Sigma_g) \Big]
+ \frac{e^{-6\lambda}}{\cos^2\zeta} d\mu^2
+ \frac{\cos^2 \zeta}{9} \widetilde  D\psi^2 \ , \nn \\
d \widetilde  D\psi &=  (2\pi) \, ( -2 \, e_2^\varphi - \chi \,  V_2^\Sigma ) \ , \qquad
\int_{S^2_\varphi} e_2^\varphi = 1 \ , \qquad
\int_{\Sigma_g} V_2^\Sigma =   1  \ , \nn \\
ds^2(S^2_\varphi)^{\rm g} & =  Dy^a \, Dy_a \ , \qquad y^a \, y_a = 1 \ , \qquad
Dy^a = dy^a - A^{ab} \, y_b \ , \qquad
A^{ab} \equiv \epsilon^{abc} \, A_c \ .
\end{align}
The 2-form $e_2^\varphi$ is defined in \eqref{e2_def}.
The gauge fields $A^{a}$ for $SO(3)_\varphi$ also enter $G_4$,
as described around \eqref{final_E4}.
We replace $V_2^\varphi$ with $e_2^\varphi$,
and 
  $\widehat D\psi$ with
$\widetilde D\psi$.  
Therefore, the form of $G_4$ we use for the
reduction is 
\begin{align}
G_4 &= 2\pi \, d \Big[  \big(\widetilde \gamma_\varphi \, e_2^\varphi
+ \widetilde \gamma_\Sigma \, V_2^\Sigma \big) \wedge \widetilde D \psi \Big] \ .
\end{align}

The dimensional reduction of the 11d Ricci scalar yields
\begin{align}
R_{(11)} & = e^{- 2 \lambda} \, R_{(5)}
- \frac 14 \, e^{-4\lambda} \, \bigg[
\frac 19 \, e^{2\lambda} \, \cos^2\zeta \, y^a \, y^b
+ e^{-4\lambda} \, F_\varphi \, (\delta^{ab} - y^a \, y^b)
\bigg]   \, F^{a}_{mn} \, F^{b \, mn} + \dots
\end{align}
where $m,n = 0,\dots,4$ are indices in external 5d spacetime,
$R_{(5)}$ is the Ricci scalar of the external 5d metric,
and we have only written down the terms that are relevant for our discussion.
We also have
\beq
*_{11} 1  = \frac 13 \, e^{-4\lambda} \, F_\varphi \, F_\Sigma \, 
(4\pi \, V_2^\varphi) \wedge (- 2\pi\chi \, V_2^\Sigma) \wedge d\mu \wedge d\psi \wedge (*_5 1) \ ,
\eeq
with $V_2^\varphi$, $V_2^\Sigma$ normalized as in \eqref{GMSW_review_metric},
and $*_5$ denoting the Hodge star with respect to the external 5d metric.
Finally, one computes
\begin{align}
G_4 \wedge *_{11} G_4 & = \frac{e^{-6\lambda } \, F_\varphi }{12 \, \chi ^2\, F_\Sigma} \,
\bigg[ 9 \,\chi^2 \, F_\Sigma^2 \, \bigg( \frac{ d\widetilde \gamma_\varphi}{d\mu} \bigg)^2
+ e^{6\lambda} \, (2 \, \widetilde \gamma_\Sigma + \chi \, \widetilde \gamma_\varphi)^2 
\bigg] \, z^a \, z^b \,    F_{a} \wedge *_5 F_{b}   \wedge{} \nn \\
& \wedge  (4\pi \, V_2^\varphi) \wedge ( - 2 \pi  \chi \, V_2^\Sigma)
\wedge d\mu \wedge d\psi    + \dots
\end{align}
where we have only written down the terms that can saturate the integration
along the internal directions.
Notice that our conventions for the Hodge star 
is such that $F^a \wedge *_5 F_a = \frac 12 \, F^a_{mn} \, F_a^{mn} \, *_5 1$.

We are now in a position to 
perform the integral over the internal directions.
The result reads
\beq \label{reduction_result}
\int_{M_{11}} \bigg[ R_{(11)} *_{11} 1
- \frac 12 \, G_4 \wedge *_{11} G_4\bigg]
= \int_{M_5 } \bigg[ 
\alpha_1 \, R_{(5)} *_{5} 1 + \alpha_2 \, F^a \wedge *_5 F_a + \dots \bigg ]  \ ,
\eeq
with the coefficients $\alpha_1$, $\alpha_2$ given by
\begin{align}
\alpha_1 & = \bigg[ 
- \frac{8}{81} \, \pi^3 \, \mu \, \chi \, \Big( 
3 \, a_\Sigma + 3 \, k \, a_\varphi + 2 \, k \, \mu^2
\Big)
  \bigg]_{-\mu_{\rm N}}^{+\mu_{\rm N}} \ ,      \nn \\
\alpha_2 & = \bigg[ 
\frac{4 \pi ^3 \mu  \chi }{2187 \left(a_\Sigma -k \mu ^2\right){}^3} \, 
\bigg(
6 k \mu ^6 a_{\Sigma }-15 k \mu ^2 a_{\Sigma } a_{\varphi }^2+k \mu ^2 a_{\Sigma }^3+3 k
   a_{\Sigma } a_{\varphi }^3+45 k a_{\Sigma }^3 a_{\varphi }
   \nn \\
   &+18 \mu ^6 a_{\varphi }+15 \mu ^4
   a_{\Sigma }^2+9 \mu ^4 a_{\varphi }^2-9 \mu ^2 a_{\Sigma }^2 a_{\varphi }-\mu ^2 a_{\varphi
   }^3+24 a_{\Sigma }^2 a_{\varphi }^2+24 a_{\Sigma }^4+8 \mu ^8
   \bigg)
  \bigg]_{-\mu_{\rm N}}^{+\mu_{\rm N}} \ .
\end{align}

We adopt the following parametrization of the 5d effective action,
\beq
S_5 = \int_{M_5} \bigg[
\frac{1}{16\pi G_N^{(5)}} \, R_{(5)} *_5 1
+ \frac{1}{g_{SG}^2} \, {\rm Tr}_{\rm f} (F \wedge *_5 F) + \dots \bigg] \ ,
\eeq
where
the trace is in the fundamental representation of $SU(2)_\varphi$,
with conventions
\beq 
{\rm Tr}_{\rm f} (T^a \, T^b) = \frac 12 \, \delta^{ab} \ , \qquad
{\rm Tr}_{\rm f} (F \wedge *_5 F) = \frac 12 \, F^a \wedge *_5 F_a  \ ,
\eeq
where $T^a$ are the generators of $SU(2)_\varphi$.
Our definition of $g^2_{SG}$ agrees with the conventions
of \cite{Freedman:1998tz}.
Keeping into account the prefactor $1/(2\kappa_{11}^2)$
in the M-theory action \eqref{Mtheory_action},
the reduction result \eqref{reduction_result} translates into
\beq
G_N^{(5)} = 2^4 \, \pi^7 \, \ell_p^{9} \, \alpha_1^{-1} \ , \qquad
g_{SG}^2 = 2^{7} \, \pi^8 \, \ell_p^9 \, \alpha_2^{-1} \ .
\eeq
The holographic central charge $c$ and $SO(3)_\varphi$ flavor
central charge $B$, in the notation of \cite{Freedman:1998tz},
are given in terms of $G_N^{(5)}$, $g_{SG}$ as
\beq
c = \frac{\pi \, L_{AdS}^3}{8 \, G_N^{(5)}} \ , \qquad
B = \frac{8\pi^2}{g_{SG}^2} \ ,
\eeq
so that we have the identifications
\beq
c = 2^{-7} \, \pi^{-6} \, \bigg[ \frac{\ell_p}{L_{AdS}} \bigg]^{-9} \, \alpha_1 \ , \qquad
B = 2^{-4}\, \pi^{-6} \, \bigg[ \frac{\ell_p}{L_{AdS}} \bigg]^{-9} \, \alpha_2 \ .
\eeq

For definiteness, we proceed in the case in which the Riemann surface
has genus $g \ge 2$ and the parameter $\gamma_0$ is set to zero.
We may then use the relations \eqref{muN_expr}, \eqref{ratioVSratio},
\eqref{ellpcube}
and express $c$ and $B$ in terms of $\tilde r$,
\begin{align}
c & = \frac{9}{32} \, \chi \, N^3 \, \bigg[
(\tilde r-1) \, (2 \, \tilde r^2 - \tilde  r + 2) - 2 \, (\tilde r^2 - \tilde r + 1 )^{3/2}
 \bigg]  \ ,    \label{sugra_c} \\
 B & = - \frac{1}{16} \, \chi \, N^3\,  \bigg[
 - \frac 83 \, (3 \, \tilde r^2 - 3 \, \tilde r + 2 ) + 8 \, (\tilde r - 1) \, \sqrt{ \tilde r^2 - \tilde r + 1 }
 \bigg]  \ .  \label{sugra_B}
\end{align}
To get the above expressions we had to de-nest some nested radicals.
The supergravity  results for $c$, $B$ match perfectly
with the large-$N$ field theory analysis performed in the main text.

\bibliographystyle{./ytphys}
\bibliography{./refs}

\end{document}